\documentclass[a4paper,fleqn,usenatbib,useAMS]{mnras}

\usepackage{hyperref}
\usepackage{graphicx}	
\usepackage{amsmath}	
\usepackage{amssymb}	
\usepackage{multicol}        
\usepackage{bm}		
\usepackage{pdflscape}	
\usepackage{color}
\usepackage[normalem]{ulem}

\newcommand{\cannon}{\textit{The Cannon}}
\newcommand{\sme}{SME}
\newcommand{\afe}{$\alpha$}
\newcommand{\gradunit}{dex kpc$^{-1}$}

\usepackage[T1]{fontenc}
\usepackage{ae,aecompl}
\usepackage{txfonts}
\usepackage{gensymb}
\usepackage{url}

\title[The GALAH Survey: Galactic disk properties]{The GALAH survey: properties of the Galactic disk(s) in the solar neighbourhood}

\author[L. Duong et al.]{L. Duong,$^{1}$\thanks{Contact e-mail: \href{mailto:ly.duong@anu.edu.au}{ly.duong@anu.edu.au}}, K. C. Freeman,$^{1}$ M. Asplund,$^{1,2}$ L. Casagrande,$^{1}$ S. Buder,$^{2}$ K. Lind,$^{2,3}$
\newauthor{M. Ness,$^{2}$ J. Bland-Hawthorn,$^{2,4}$ G. M. De Silva,$^{4,6}$ V. D'Orazi,$^{7}$ J. Kos,$^{4}$} G. F. Lewis,$^{4}$
\newauthor{J. Lin,$^{1}$ S. L. Martell,$^{8}$ K. Schlesinger,$^{1}$ S. Sharma,$^{4}$ J. D. Simpson,$^{6}$ D. B. Zucker,$^{9}$ }
\newauthor{T. Zwitter,$^{10}$ B. Anguiano,$^{9,11}$ G. S. Da Costa,$^{1}$ E. Hyde,$^{12}$ J. Horner,$^{13}$ P. R. Kafle,$^{14}$}
\newauthor{D. M. Nataf,$^{1,15}$ W. Reid,$^{16,17}$ D. Stello,$^{4,8,18}$ Y.-S. Ting$^{1,19,20,21}$} and R. F. G. Wyse$^{15}$
\\
$^{1}$Research School of Astronomy \& Astrophysics, Australian National University, ACT 2611, Australia\\
$^{2}$Max-Planck-Institute for Astronomy, Koenigstuhl 17, D-69117 Heidelberg, Germany\\
$^{3}$Department of Physics and Astronomy, Uppsala University, Box 516, SE-751 20 Uppsala, Sweden\\
$^{4}$Sydney Institute for Astronomy, School of Physics A28, University of Sydney, NSW 2006, Australia\\
$^{6}$Australian Astronomical Observatory, North Ryde, NSW 1670, Australia\\
$^{7}$INAF-Osservatorio Astronomico di Padova, Vicolo dell'Osservatorio 5, 35122 Padova, Italy\\
$^{8}$School of Physics, University of New South Wales, NSW 2052, Australia\\
$^{9}$Department of Physics and Astronomy, Macquarie University, NSW 2109 Australia\\
$^{10}$Faculty of Mathematics and Physics, University of Ljubljana, Jadranska 19, 1000 Ljubljana, Slovenia\\
$^{11}$Department of Astronomy, University of Virginia, P.O. Box 400325 Charlottesville, VA 22904-4325, USA\\
$^{12}$Western Sydney University, Locked Bag 1797,  Penrith South DC, NSW 1797, Australia\\
$^{13}$University of Southern Queensland, Toowoomba, Queensland 4350, Australia\\
$^{14}$ICRAR, The University of Western Australia, 35 Stirling Highway, Crawley WA 6009, Australia\\
$^{15}$Center for Astrophysical Sciences and Department of Physics and Astronomy, The Johns Hopkins University, Baltimore, MD 21218, USA\\
$^{16}$Department of Physics and Astronomy, Macquarie University, Sydney, NSW 2109, Australia\\
$^{17}$Western Sydney University, Locked bag 1797, Penrith South DC, NSW 2751, Australia\\
$^{18}$Department of Physics and Astronomy, Stellar Astrophysics Centre, Aarhus University, DK-8000 Aarhus C, Denmark\\
$^{19}$Institute for Advanced Study, Princeton, NJ 08540, USA\\
$^{20}$Department of Astrophysical Sciences, Princeton University, Princeton, NJ 08544, USA \\
$^{21}$Observatories of the Carnegie Institution of Washington, Pasadena, CA 91101, USA
}

\date{Accepted XXX. Received YYY; in original form ZZZ}

\pubyear{2018}

\begin{document}
\label{firstpage}
\pagerange{\pageref{firstpage}--\pageref{lastpage}}
\maketitle

\begin{abstract}
Using data from the GALAH pilot survey, we determine properties of the Galactic thin and thick disks near the solar neighbourhood. The data cover a small range of Galactocentric radius ($7.9 \la R_\mathrm{GC} \la 9.5$ kpc), but extend up to 4 kpc in height from the Galactic plane, and several kpc in the direction of Galactic anti-rotation {(at longitude $260 ^\circ \leq \ell \leq 280^\circ$)}. This allows us to reliably measure the vertical density and abundance profiles of the chemically and kinematically defined `thick' and `thin' disks of the Galaxy. The thin disk (low-\afe~population) exhibits a steep negative vertical metallicity gradient, at d[M/H]/d$z=-0.18 \pm 0.01$ \gradunit, which is broadly consistent with previous studies. In contrast, its vertical \afe-abundance profile is almost flat, with a gradient of d[\afe/M]/d$z$ = $0.008 \pm 0.002$ \gradunit. {The steep vertical metallicity gradient of the low-\afe~population is in agreement with models where radial migration has a major role in the evolution of the thin disk}. The thick disk (high-\afe\ population) has a weaker vertical metallicity gradient d[M/H]/d$z = -0.058 \pm 0.003$ \gradunit. The \afe-abundance of the thick disk is nearly constant with height, d[\afe/M]/d$z$ = $0.007 \pm 0.002$ \gradunit. The negative gradient in metallicity and the small gradient in [\afe/M] indicate that the high-\afe~population experienced a settling phase, but also formed prior to the onset of major SNIa enrichment. We explore the implications of the distinct \afe-enrichments and narrow [\afe/M] range of the sub-populations in the context of thick disk formation. 
\end{abstract}




\section{Introduction}  
\label{intro}
The Milky Way is believed to have a thick disk, similar to those observed photometrically in external disk galaxies ~\citep{Tsikoudi1979,Burstein1979a,Dalcanton2002,Yoachim2006,Comeron2015}.  The ubiquity of thick disks indicates that they are an integral part of disk galaxy evolution. The Galactic thick disk was originally discussed as a distinct structural component by~\cite{gr83}\footnote{See also~\cite{Yoshii1982}.}, who showed that the vertical stellar density profile at the Galactic South pole was best described by two exponentials. Much debate has since ensued over the origin and properties of the Galactic thick disk. Most notably some authors have argued that it may not be a discrete component~\citep{Norris1987,Nemec1993,Schonrich2009,Bovy2012a}.

The chemical properties of the local thick disk have been well characterised by multiple spectroscopic studies. The consensus is that it is older (e.g., \citealt{Wyse1988,Haywood2013,Bensby2014}), kinematically hotter~\citep{Chiba2000}, and more metal-poor and \afe-rich than the thin disk~\citep{pro00,Fuhrmann2008,Bensby2014,Fuhrmann2016}. The enhanced \afe-abundances indicates that thick disk stars were enriched by SNe Type II over a short period of time, before SNe Type Ia contribution of iron-peak elements took effect in earnest. The thick disk is thought to have formed within $\approx$1--3 Gyr~\citep{Gratton2000,Mashonkina2003}, although~\citet{Haywood2013} suggested a slightly longer formation timescale of 4--5 Gyr. 

At the solar annulus, many authors have observed a gap between thin and thick disk stars in the \afe-abundance ([\afe/M]) vs metallicity ([M/H]) plane. This is widely interpreted as evidence that the thick disk is a distinct component. In recent literature, the `thick disk' is often defined chemically as the \afe-enhanced population. Large scale abundance maps from the APOGEE survey show that two distinct sequences in [\afe/M] vs [M/H] are observed at all galactocentric radii, although the fractions of stars in the two sequences varies greatly with position in the Galaxy.
In the inner Galaxy ($3<R_\mathrm{GC}<5$ kpc), and at large heights above the Galactic plane, the high-\afe~sequence dominates. Beyond galactocentric radius $R_\mathrm{GC} \approx 9$ kpc, its density decreases significantly~\citep{Hayden2015}. This observation is in line with the short scale length of about $2$ kpc for the chemical thick disk~\citep{Bensby2011,Cheng2012,Bovy2012b,Bovy2016}. The concentration of the older, \afe-enhanced population to the inner disk also indicates that the thick disk formed inside-out~\citep{Matteucci1989,Burkert1992,Samland2003,Bird2013}.
In contrast to the chemically defined thick disk of the Milky Way, the photometrically defined thick disks of external galaxies are more extended, with scale lengths comparable to thin disk scale lengths~(e.g., \citealt{Yoachim2006, Ibata2009}). {Similarly, selecting Milky Way thick disk stars using non-chemical criteria leads to a thick disk with scale length longer than that of the thin disk (e.g.,~\citealt{Ojha2001}).}

While its scale length is fairly well constrained, the scale height of the thick disk is still contentious (see \citealt{Bland-Hawthorn2016}, and references therein).~\cite{gr83} estimated the thick disk exponential scale height to be 1.35 kpc from star counts, similar to measurements made by photometric decomposition of Milky Way analogues (e.g., \citealt{Ibata2009}). More recent estimates find the thick disk scale height to be significantly shorter, and there is still some scatter in the measurements~\citep{Juric2008,Kordopatis2011,Bovy2012b,Bovy2016}. Furthermore, results from high-resolution spectroscopic surveys have raised doubts on the existence of a structurally distinct thick disk, even if there are clearly two populations with distinct \afe-enhancements.~\cite{Bovy2016} finds a smooth transition in scale-heights for mono-abundance populations, as does~\cite{Mackereth2017} for mono-age populations, where more \afe-enhanced and older stars populate increasingly greater heights.~\cite{Martig2016b} also showed that, due to flaring of the disk~{\citep{Rahimi2014,Minchev2015,Kawata2017}}, the geometrically thick part of the disk has a large age dispersion, whereas the chemical `thick disk' (high-\afe~population) has a narrow age range. This may also explain why the chemically defined thick disk of the Milky Way has a short scale length, while surface brightness measurements of geometrical thick disks in external galaxies indicate that they are radially much more extended. 

Several theoretical models have been proposed for thick disk formation and explain its observed properties. Thick disks may arise from external heating processes such as dwarf satellite accretion~\citep{aba03} or minor merger events~\citep{Quinn1986,Quinn1993,Kazantzidis2008,Villalobos2008}. The fast internal evolution of gravitationally unstable clumpy disks at high red-shift~\citep{Bournaud2009,Forbes2012} or gas-rich mergers at high red-shift~\citep{Brook2004,Brook2005} could form a thick disk. The turbulent interstellar medium observed in disk galaxies at high red-shift may also be associated with thick
disk formation~(e.g.\citealt{Wisnioski2015}). Radial migration of stars~\citep{selbin02}, where stars are transported outwards and gain vertical height to form a thick disk, is another possibility that has been extensively discussed~\citep{Schonrich2009,Minchev2010,Loebman2011,Roskar2012,Schonrich2017}. Although there is evidence for radial migration
in the thin disk, such as the presence of very metal rich low-\afe~stars in the solar neighbourhood~{\citep{Haywood2008,Casagrande2011}} and the skewness of metallicity distribution functions at different Galactic radii~\citep{Hayden2015,Loebman2016}, the role of radial migration in thick disk formation is still unclear, and is not supported by some observed properties of the thick disk (high-\afe) population (e.g., \citealt{Haywood2013,Recio-Blanco2014,Bovy2016}). \cite{Vera-Ciro2014} showed in their simulation that radial migration can have strong effects on the thin disk, but not the thick disk.{~\cite{Aumer2016} found that in their standard model, outwardly-migrating stars are not responsible for the creation of the thick disk, but thick disks can form in models with high baryon fractions. However, in their high-baryon models, the bar is too long, the young stars are too hot and the disk is strongly flared}. 

Observational evidence to discern thick disk formation scenarios is still inconclusive.
Earlier results, such as the lack of thick disk vertical metallicity gradient observed by~\cite{Gilmore1995} and orbital eccentricity distributions by~\cite{Sales2009} and~\cite{Dierickx2010}, favoured merger scenarios. More recent studies, most of which separate thin and thick disk stars by their metallicity or kinematics, indicate that the thick disk does have a vertical metallicity gradient~\citep{Chen2011,Katz2011,Kordopatis2011,Ruchti2011}, but the gradients measured by these studies vary greatly due to their different methods of isolating the thick disk. Few studies report on the vertical abundance profile of the disk, although an accurate measurement of the metallicity and abundance profile as a function of distance from the Galactic plane can provide important constraints for the evolution history of the disk.

This work is motivated by the current uncertainty about the formation and properties of the Galactic thick disk. The thick disk is important because its formation is a seemingly ubiquitous feature of disk galaxy evolution; its rapid formation and old population means that it provides a detailed snap-shot of the conditions in the early Galaxy. Understanding how the thick disk formed and evolved will be central to chemical tagging efforts of current and future high resolution massive spectroscopic surveys such as 4MOST~\citep{DeJong2011}, \textit{Gaia}-ESO~\citep{Gilmore2012}, APOGEE~\citep{Majewski2015}, GALAH\footnote{\url{www.galah-survey.org}}~\citep{DeSilva2015}, and WEAVE~\citep{Dalton2016}.

We have used data from the first GALAH survey internal release to study the properties of the Galactic thick disk. We show that at the solar circle, the thick disk exhibits a non-negligible vertical metallicity gradient, and the thin disk shows a steep vertical metallicity gradient.
We find that the mean \afe-element abundance does not vary significantly with height in either of the chemically and kinematically defined thick and thin disks.

The paper is structured as follows: Section \ref{sample} describes the stellar sample used in the analysis, including field and colour selection. Section \ref{analysis} explains the methods of obtaining stellar parameters, abundances and the distances, as well as how thin and thick disk components were defined. Section \ref{bias} explores the possible effects of our selection and how they were corrected for. Section \ref{gradients} presents the results of metallicity and the \afe-abundance variation with vertical height are described in section \ref{profiles}. We discuss the implications of our results for the formation and evolution of the thick disk in Section \ref{discussion}, and summarise the work in Section \ref{conclusions}.

\section{Sample selection} 
\label{sample}

\begin{figure*}
	\includegraphics[width=0.95\columnwidth,trim={0.5cm 0.5cm 0.5cm 0.5cm},clip]{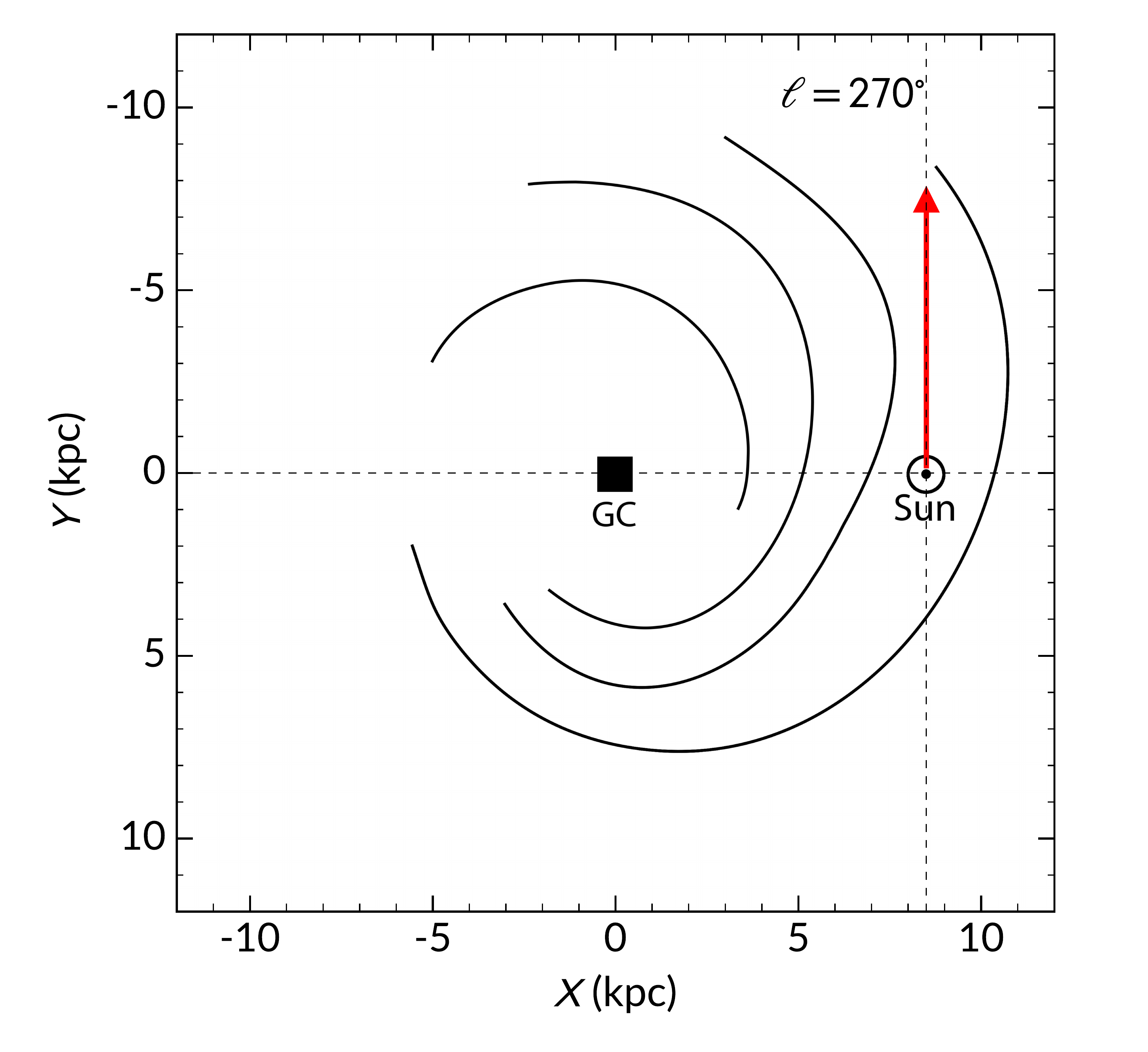} \includegraphics[width=0.975\columnwidth,trim={0.5cm 0.5cm 0.5cm 0.5cm},clip]{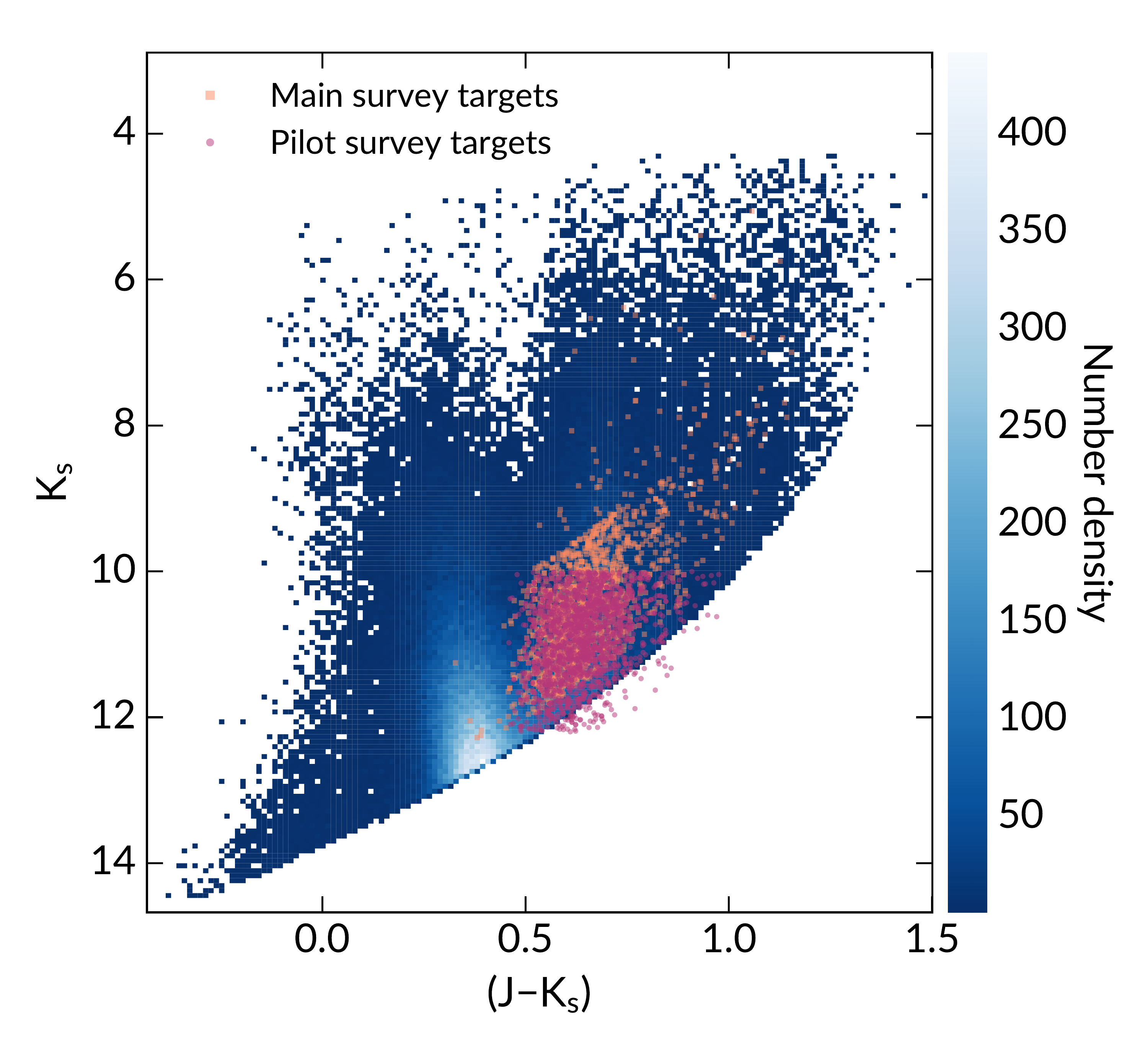}
	\caption{The location of observed fields and colour-magnitude selection of the GALAH main and pilot survey targets. Left panel: a schematic view of the Milky Way, illustrating the targeted line-of-sight. We observe along $\ell \approx 270^\circ$ and at five latitudes below the plane. The spiral arms are shown as traced by \ion{H}{II} gas, from~\protect\cite{Drimmel2001}. Right panel: The colour-magnitude selection of the  stars in this analysis is shown against all GALAH input catalogue targets within the observed region (including special bright targets). The pilot survey has a simple magnitude cut, at bright and faint limits of $K_s= 10$ and $12$, respectively. The main GALAH survey magnitude selection $12 < V_{JK} < 14$ appears as a stripe in the ($J-K_{s}$) vs $K_{s}$ plane. The pilot survey extends slightly fainter than the main survey, and the handful of stars falling outside of the main survey selection are from a bright field.}
	\label{fig:field_loc} 
\end{figure*}

\begin{figure*}
	\includegraphics[width=1.2\columnwidth,trim={1cm 1cm 0.5cm 0cm},clip]{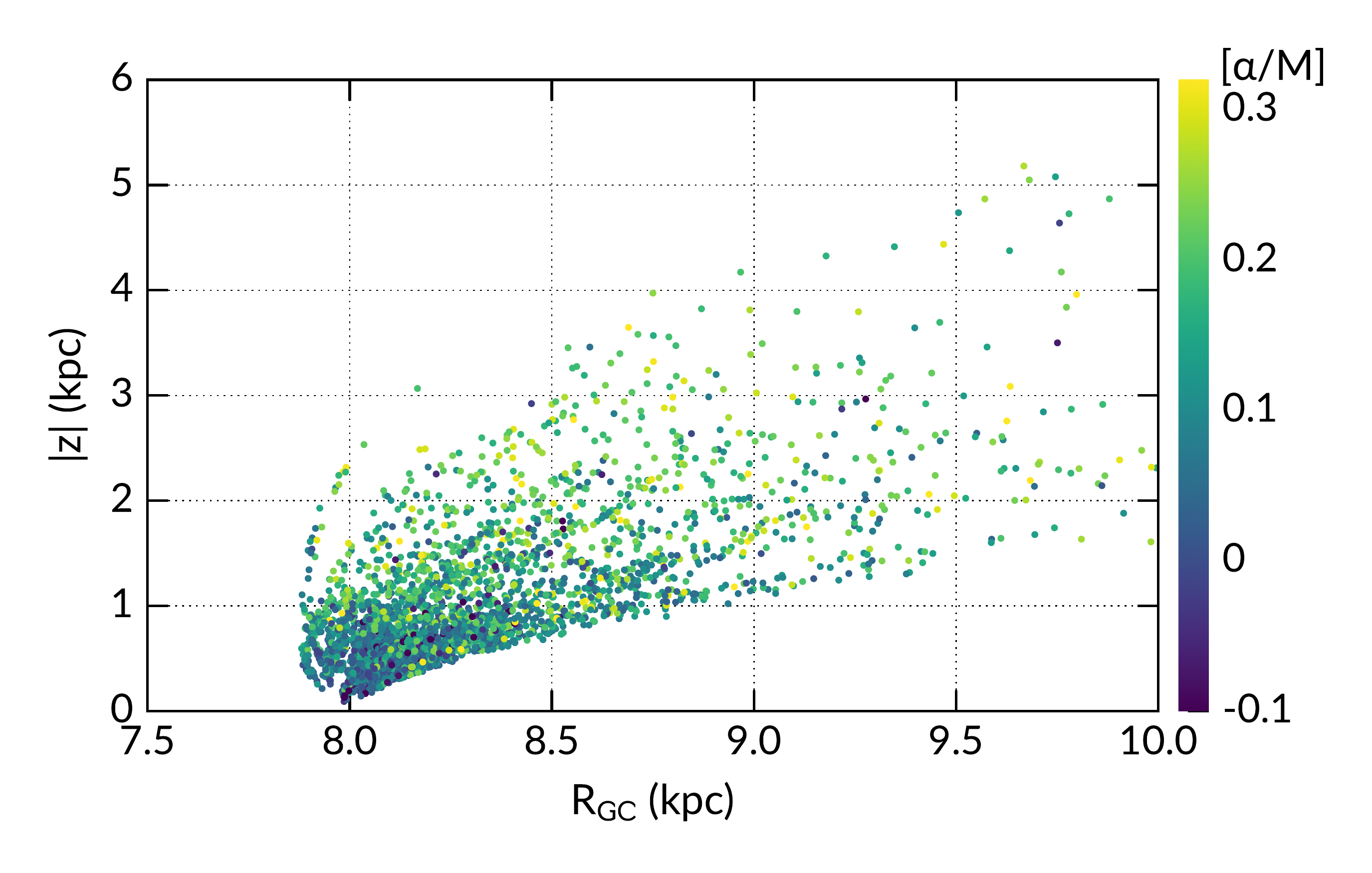}
	\caption{The distribution in Galactocentric radius and height below the plane for the entire sample, adopting \textit{R}\textsubscript{GC,}$_{\odot}$ = 8 kpc. There are a few stars (not shown) outside the limits |\textit{z}| $> 6$ kpc and \textit{R} $> 10$ kpc. Stars are colour-coded by $\alpha$-abundances: low-\afe~stars lie typically closer to the plane whereas \afe-enhanced stars are found at greater distances from the plane.}
	\label{fig:Rz}
\end{figure*}

We present in this paper results from the GALAH survey internal data release v1.3. The data acquisition and reduction are described in~\cite{Martell2017} and~\cite{Kos2017}, respectively. The stellar parameter and abundance determination is summarised in Section \ref{spp}. Briefly, GALAH spectra cover four optical bands, at wavelengths located within the Johnson-Cousins $B,V,R,I$ passbands, with resolving power $\lambda/\Delta \lambda \approx 28 000$~\citep{DeSilva2015}. The GALAH main-survey selects stars according to a simple magnitude criterion: $12<V_{JK}<14$, where the $V_{JK}$ magnitude is estimated from 2MASS~\citep{Skrutskie2006} $J,K_s$ photometry via the transformation:
\begin{equation}
V_{JK}=K_s +2(J-K_s +0.14)+0.382 \,e^{\,(J-K_s-0.2)/0.50}
\label{e1}
\end{equation}
\noindent The above equation is discussed further in \cite{Sharma2018} (see their Fig. 1). The magnitude selection in $V_{JK}$ manifests as a $(J-K_s)$ colour dependence when plotted as function of other magnitudes, as shown in Fig. \ref{fig:field_loc}, right panel. 

In addition to normal survey fields which follow the $V_{JK}$ magnitude limit described above, GALAH also observed special fields, such as pilot survey fields (which included benchmark stars and clusters), and bright stars selected from the $Tycho$-2 catalogue~\citep{Martell2017}, most of which also appear
in the {\textit Gaia} DR1 catalog~{\citep{Brown2016}}.

As part of the GALAH pilot survey, we conducted a study of the chemical properties and distribution of the Galactic thin and thick disks. Fields were chosen towards Galactic longitude $\ell=270^{\circ}$, as shown in Fig.~\ref{fig:field_loc}, left panel. This longitude was chosen to maximise the asymmetric drift component between thin and thick disk stars \citep{Gilmore2002,Wyse2006}, thus making it easier to distinguish them by their kinematics. We observed fields at five latitudes:  $b=-16^{\circ},-22^{\circ},-28^{\circ},-34^{\circ}$ and $-42^{\circ}$. Fig.~\ref{fig:Rz} shows the distribution of observed stars in Galactic coordinates $R_{\mathrm{GC}}$ and $|z|$ (distances are derived as per Section \ref{dist}). Adopting $R_{\mathrm{GC},\odot} = 8$ kpc, most of the stars are concentrated around the solar radius, between $R_{\mathrm{GC}} = 8-8.5$ kpc, and up to about 4 kpc in height below the Galactic plane. {Since our longitude range is between $\ell=(260,280)$, we also observed stars with $R_{\mathrm{GC}} < 8$ kpc.}

We chose to use only giants in this study to include a larger range of distances and heights from the plane. The magnitude limits of the main GALAH survey result in giants making up only about $25$\% of stars observed. In order to increase the fraction of giants, a colour cut at $(J-K_s) > 0.45$ was imposed for the pilot survey prior to observations, which excludes turn-off stars and some dwarfs. We also extended the faint $V_{JK}$-magnitude limit of the pilot survey to 14.5 in order to observe a larger fraction of clump giants. Also included in this analysis are giants from the GALAH main-survey that fall within the same Galactic longitude--latitude range described above. The colour and magnitude selection for all stars included in the analysis is shown in Fig. \ref{fig:field_loc}.

\section{Data analysis} 
\label{analysis}

\subsection{Stellar parameters and alpha abundances}
\label{spp}

The GALAH stellar parameters and abundances pipeline will be described in detail elsewhere; here we seek to give a brief summary. The pipeline is a two-step process, involving spectral synthesis using \sme~(\textit{Spectroscopy Made Easy}) ~\citep{Valenti1996,Piskunov2017} and the data-driven generative modelling approach of \cannon~\citep{Ness2015}. We identify a sample of stars with high signal-to-noise ratio, each visually inspected to be free of irregularities like unexpected continuum variations and large cosmic ray residuals. This set of stars serves as the training set, the labels of which are propagated to all other survey stars. The training set includes $Gaia$ benchmark standards~\citep{Jofre2014,Heiter2015} whose parameters have been determined by non-spectroscopic methods; globular and open clusters and stars with accurate asteroseismic surface gravity from K2 Campaign 1~\citep{Stello2017}. In total there are $\approx$2500 training stars.

In the first step, stellar parameters for the training set are obtained with \sme. Here we use the \textsc{{marcs}} model atmospheres~\citep{Gustafsson2008}, and non-LTE corrections for Fe~\citep{Lind2012}. \sme~syntheses of H\afe~and H$\beta$, neutral and ionised lines of Ti, Sc and Fe are used to determine $T_\textrm{eff}$, $\log g$, [M/H]\footnote{We use [M/H] to denote metallicity to differentiate it from the actual iron abundance [Fe/H]. The metallicity reported in this data release is the iron abundance of the best-fit atmospheric model and mostly measured from Fe lines. However, [M/H] values are close to the true iron abundances, and GALAH results presented elsewhere have used [Fe/H] to denote metallicity, which is equivalent to the [M/H] used here.}, v$_{mic}$ (micro turbulence) and $V\sin i$ (rotational velocity), converging at the global minimum $\chi^2$. The stellar parameters are fixed when individual abundances are computed for the 
$\alpha$-elements Mg, Si, Ti. The weighted average of these elements gives [$\alpha$/M], and all abundances are scaled according to the Solar chemical composition of~\cite{Grevesse2007}. 

Although the GALAH wavelength range includes lines of the \afe-elements Ca and O, they are currently omitted from the weighted average because the Ca lines fall within problematic spectral regions (due to bad CCD pixels and/or sub-optimal data reduction), and the \ion{O}{I} triplet at 7772--7775 \AA\ is subjected to large non-LTE effects~\citep{Amarsi2015,Amarsi2016}, which are not yet accounted for in the GALAH analysis pipeline. Relative to $Gaia$ benchmark standards, \sme~produces accurate results with offsets in $\log g$ and [M/H] of $-0.15$ and $-0.1$ dex respectively, in the sense that it underestimates these values. The same surface gravity offset is also observed when SME results are compared to asteroseismic $\log g$ obtained with oscillations from \cite{Stello2017}. The offsets are constant across the HR diagram, and are corrected by simply adding 0.15 and 0.1 dex to all $\log g$ and [M/H] values of the training set prior to parameter propagation with \cannon~{\citep{Sharma2018}}. 

\begin{figure}
\centering
		\includegraphics[width=1.00\columnwidth,trim={0.5cm 0.5cm 0.5cm 0.9cm},clip]{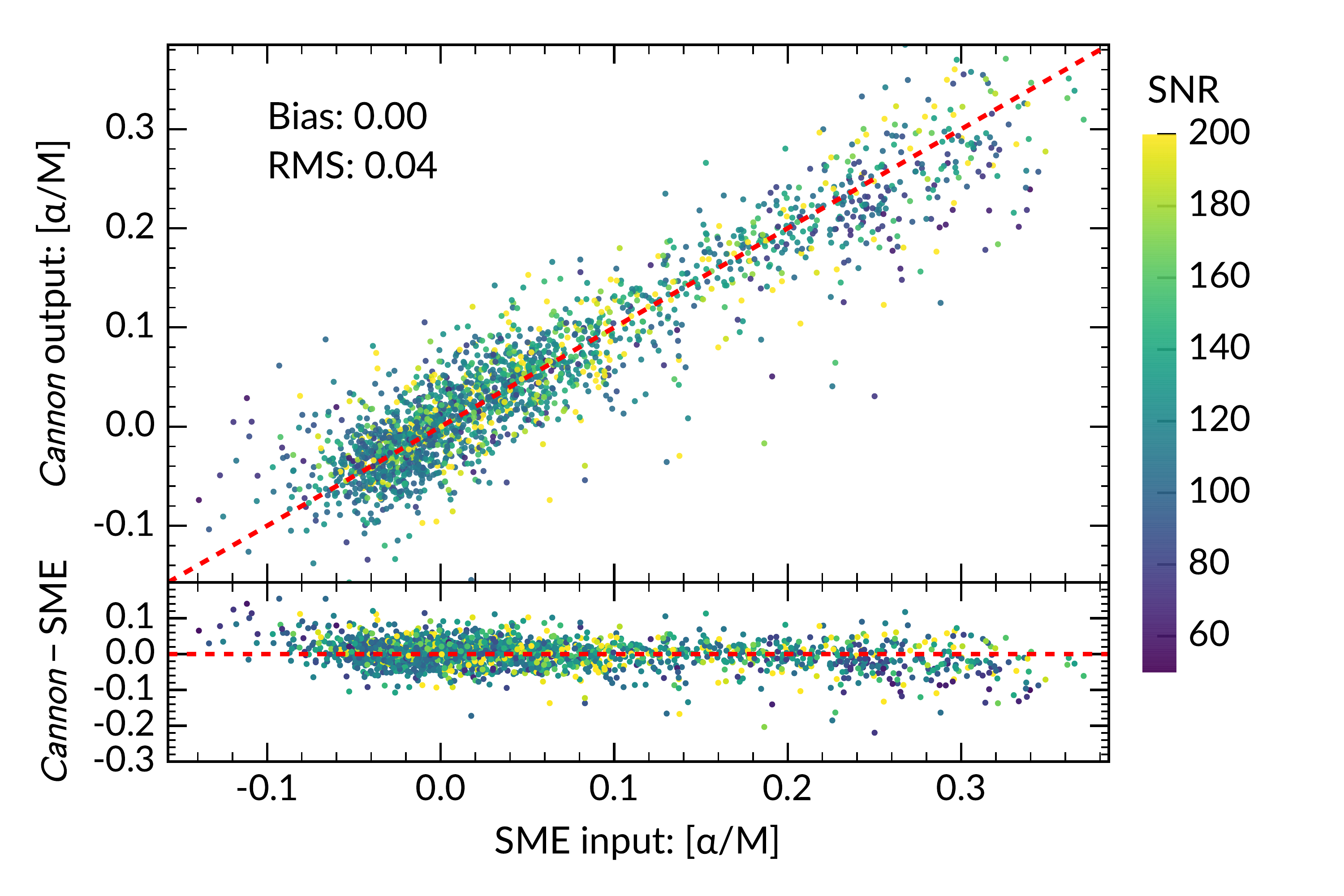}
	\caption{Cross-validation of \cannon-output against SME input for the label [\afe/M]. Stars are colour coded by signal-to-noise ratios in the `green' arm of HERMES, the wavelength of which lies within the Johnson-Cousins \textit{V}-band. The top panel shows the 1:1 relation between SME input and \cannon~output. The bottom panel shows the distribution of the difference. \cannon~reproduces the SME input without bias, and to 0.04 dex precision.}
	\label{fig:afe_precision}
\end{figure}

In the second step, \cannon~learns the training set parameters and abundances (\textit{labels}) from \sme, and builds a quadratic model at each pixel of the normalised spectrum\footnote{The normalisation method is described in \cite{Kos2017}.} as a function of the labels~\citep{Ness2015}. This model is then used to determine stellar parameters and abundances for all other survey spectra. In addition to the six primary labels described above, \cannon~uses a seventh label, extinction $A(K_s)$, to minimise the effect of reddening and diffuse interstellar bands on [\afe/M] determination. The extinction for each star of the training set was estimated with the \textsc{RJCE} method \citep{Majewski2011}. We used 2MASS $H$-band and WISE 4.5 $\mu$m photometry \citep{Wright2010}, following procedures outlined in \cite{Zasowski2013}. Parameter errors are estimated by cross-validating the input (SME) and output labels (\cannon) for the training set. Cross-validation was done by partitioning the reference set into unique sub-samples, each consisting 20\% of the full set. Five tests were performed, each time a 20\% sub-sample is left out of the training step, and used only to validate the results. Fig. \ref{fig:afe_precision} shows the combined cross-validation outcomes of all five tests for label [\afe/M]. The training set results have also been successfully applied to the TESS-HERMES survey, and the error estimation of stellar labels except for [\afe/M] is shown in Fig. 5 of~\cite{Sharma2018}. Overall,~\cannon~achieves internal precisions of 47 K in $T_\textrm{eff}$, 0.13 dex in $\log g$, 0.05 dex in [M/H] and 0.04 dex in [\afe/M], which are typical of the errors reported in this data release. 

\begin{figure}
\centering
		\includegraphics[width=0.9\columnwidth]{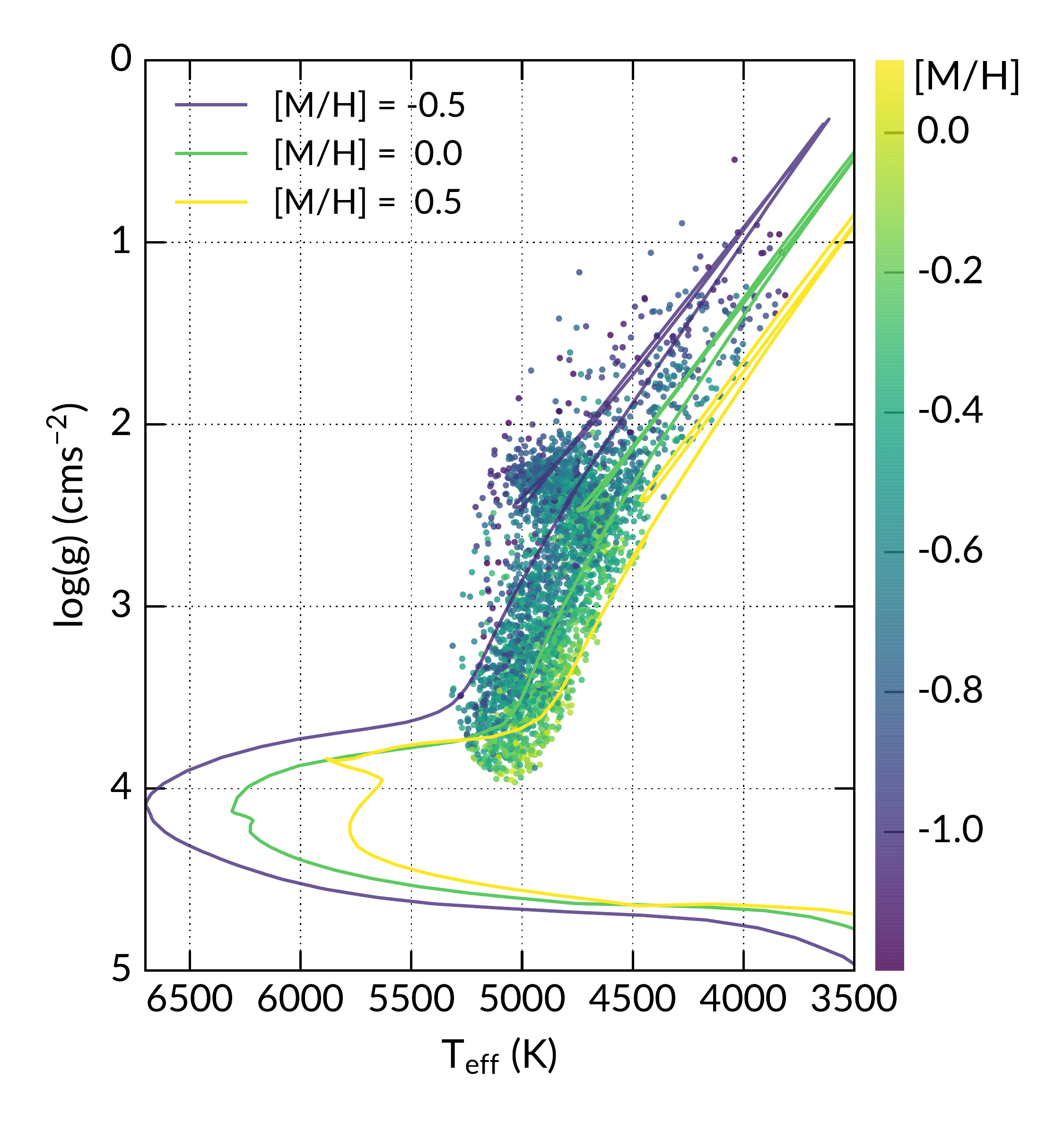}
	\caption{The \textit{Kiel} diagram from \cannon~for the sample of stars in this analysis, colour coded by [M/H]. Over plotted are 4 Gyr \textsc{parsec} isochrones at metallicities indicated in the figure legend. The stellar parameters behave as predicted by the evolutionary tracks after bias corrections to \cannon~training set (see text for details).}
	\label{fig:Cannon-HRD}
\end{figure}

\begin{figure}
\centering
		\includegraphics[width=1\columnwidth,trim={1cm 1cm 0cm 0cm},clip]{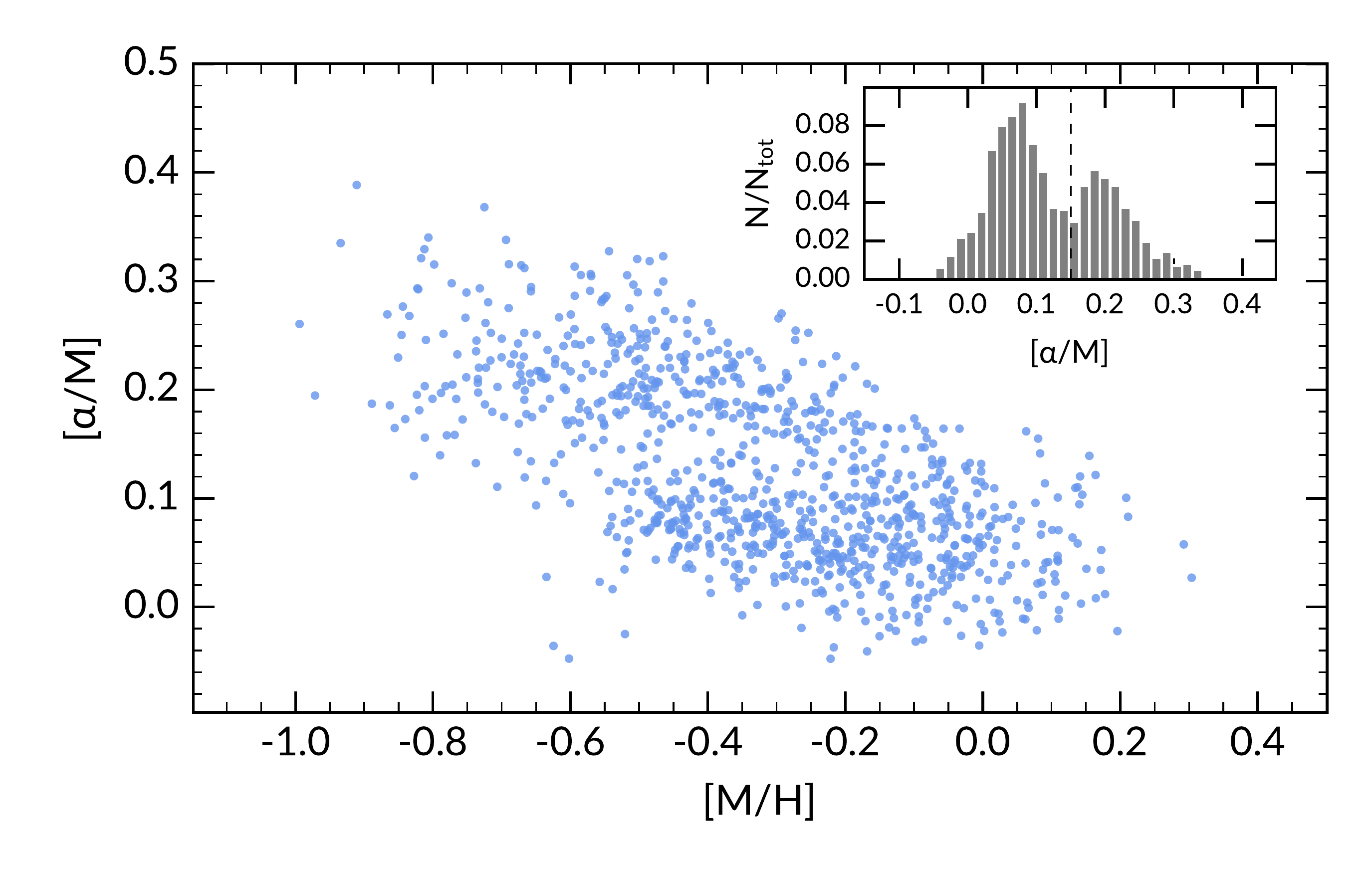}

	\caption{\afe-abundances as a function of metallicity for a sub-sample with signal-to-noise $\geq$ 80 per resolution element. There are two distinct abundance sequences corresponding to the thin disk (low-\afe) and the thick disk (high-\afe). Inset: histogram of the [\afe/M] distribution, the dotted line indicate [\afe/M] = 0.15, where the two populations appear to separate.}
	\label{fig:alpha}
\end{figure}

Fig.~\ref{fig:Cannon-HRD} shows \cannon-derived stellar parameters for the full sample of giants selected for analysis. We have excluded most sub-giants, turn-off and main-sequence stars. \cannon~is able to reproduce the accuracy and precision of \sme~such that all parameters follow the \textsc{parsec} isochrone tracks \citep{Marigo2017} without further calibrations. The [\afe/M]--[M/H] plot is shown in Fig.~\ref{fig:alpha} for a sub-sample of stars with signal-to-noise ratio $\geq$ 80 per resolution element. We observe the two distinct \afe-tracks in the [\afe/M]--[M/H] plane: a low-\afe~track extending from [M/H] $\approx$ 0.4 to $-0.6$, usually defined as the chemical thin disk, and a high-\afe~track extending from [M/H] $\approx -0.2$ to $-1$, usually defined as the chemical thick disk. The typical precision of the [\afe/M] measurements is 0.04 dex, similar to that of [M/H].

 We do not include stars with [M/H] $\leq -1$ dex here, because few metal-poor stars could be used in the training set (the stars are rare, and typically have low SNR), rendering \cannon~results for metal-poor stars significantly less accurate. \cannon~has limited ability to extrapolate, which is evident in the comparison to $Gaia$ benchmarks: stars with [M/H] $< -1$ have larger deviations from reference values~\citep{Sharma2018}. {This does, however, exclude the metal-weak thick disk from our analysis. We find that other studies which include the metal-poor extension of the thick disk, such as~\cite{Katz2011, Ruchti2011} reported similar results to ours (see detailed discussion in Sections \ref{subsubsec:highalphafeh} and \ref{profiles}). Furthermore, there are few stars with [M/H] $\leq -1$ dex to begin with (2\% of the full sample), so their exclusion may have small effects on the vertical gradients derived in later sections, but it is unlikely that this would have a major impact on our conclusions.}

\subsection{Distance determination}
\label{dist}
Distances are typically determined by isochrone fitting methods using the fundamental stellar parameters $T_\textrm{eff}$, $\log g$ and [M/H] and photometry. Theoretical constraints, such as stellar evolution and initial mass functions (IMF) have been included by \cite{Zwitter2010} and \cite{Burnett2010}, respectively, to obtain more accurate distances for the RAVE survey \citep{Steinmetz2006}. Isochrone distances are dependent on all fundamental parameters, but a strong dependence on [M/H] can cause correlated errors when trying to assess the metallicity distribution as a function of distance from the Galactic centre or above the plane \citep{Schlesinger2014,Anguiano2015}. In this section we describe an empirical method of distance determination that does not have such a strong dependence on inferred [M/H], which may be advantageous for our measurements of metallicity vertical gradients in Section \ref{gradients}.

\begin{figure}
\includegraphics[width=1.03\columnwidth,trim={=0.8cm 1cm 0cm 0cm},clip]{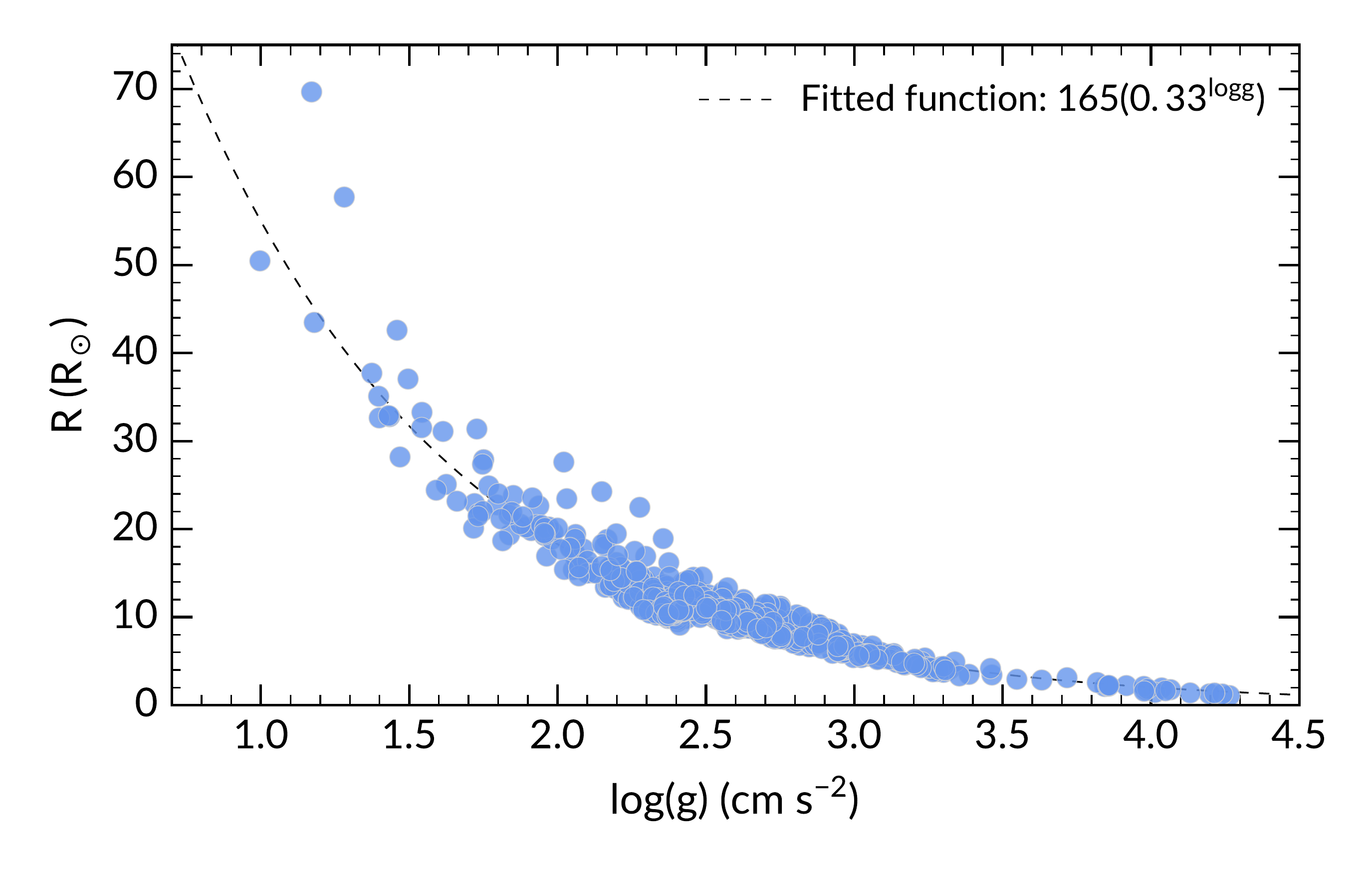}
\caption{The correlation between stellar surface gravity and radius, data from the \textit{Kepler} sample of \protect\cite{Casagrande2014}. The stellar radius as a function of $\log g$ is best described by an exponential.}
\label{fig:dist-poly}
\end{figure}

To determine the distance, we exploit the relationship between stellar surface gravity and radius $R$ using \textit{Kepler} asteroseismic data from~\cite{Casagrande2014}. Fig. \ref{fig:dist-poly} shows the $\log g-R$ correlation and the exponential function that best fits the data. Using spectroscopically determined $\log g$, we compute for each GALAH star a radius (in solar radii) using the function:
\begin{equation}
R_* = 165 (0.33^{\log g})
\end{equation}
\noindent Note that starting from the definition of $ g = GM/R_*^2$, where $G$ is the gravitational constant, and $M$ the mass of the star, we arrive at the formula:
\begin{equation}
R_*/R_\odot = 10^{0.5 (\log g_\odot - \log g)},
\label{eqn:dist}
\end{equation}
\noindent which is equivalent to $165.59 * 0.316^{\log g}$, assuming that $M = M_\odot$ and the solar $\log g$ is 4.438068 cm $\mathrm{s}^{-2}$ . However, the function used to fit the data in Fig \ref{fig:dist-poly} returns the minimum reduced-$\chi^2$ (perhaps due to differences in the stellar mass compared to the assumed solar value), and is used instead\footnote{ The typical difference between distances derived using $165.59 * 0.316^{\log g}$ and equation \ref{eqn:dist} is 13\%, which is comparable to the distance uncertainties (Section \ref{disterr}).}.

\noindent The absolute luminosity is estimated using the effective temperature and radius relation:
\begin{equation}
L=4 \pi R_*^2 \sigma T_\textrm{eff}^4
\end{equation}

\noindent {Finally, we interpolate the stellar parameters $T_\textrm{eff}$, $\log g$, [M/H] over a grid of synthetic spectra to determine the correction that needs to be applied to 2MASS $J, H, K_s$ photometry to derive the bolometric flux $\mathcal{F}_{bol}$\footnote{Although [M/H] is used, the dependence of $\mathcal{F}_{bol}$ on this parameter is minimal~\citep{Casagrande2010}}. We correct for extinction using the~\cite{Schlegel1998} map to de-redden the observed 2MASS magnitudes. This is done using extinction coefficients computed on-the-fly for the set of stellar parameters adopted~\citep{Casagrande2010}.} The distance is then simply:
\begin{equation}
D=\left(\frac{L}{4 \pi \mathcal{F}_{bol}}\right)^{1/2}
\end{equation}

\noindent As is evident from Fig. \ref{fig:dist-poly}, the $\log g$--stellar radius relation is poorly constrained for $\log g \leq 1.5$, because we have few seismic data points in this region and the scatter is larger. There are however  relatively few stars with $\log g < 1.5$ in our sample (see Fig. \ref{fig:Cannon-HRD}).

\subsubsection{Distance error estimate}
\label{disterr}
\begin{figure*}
\centering
		\includegraphics[width=1.2\columnwidth,trim={0cm 1cm 0cm 0cm},clip]{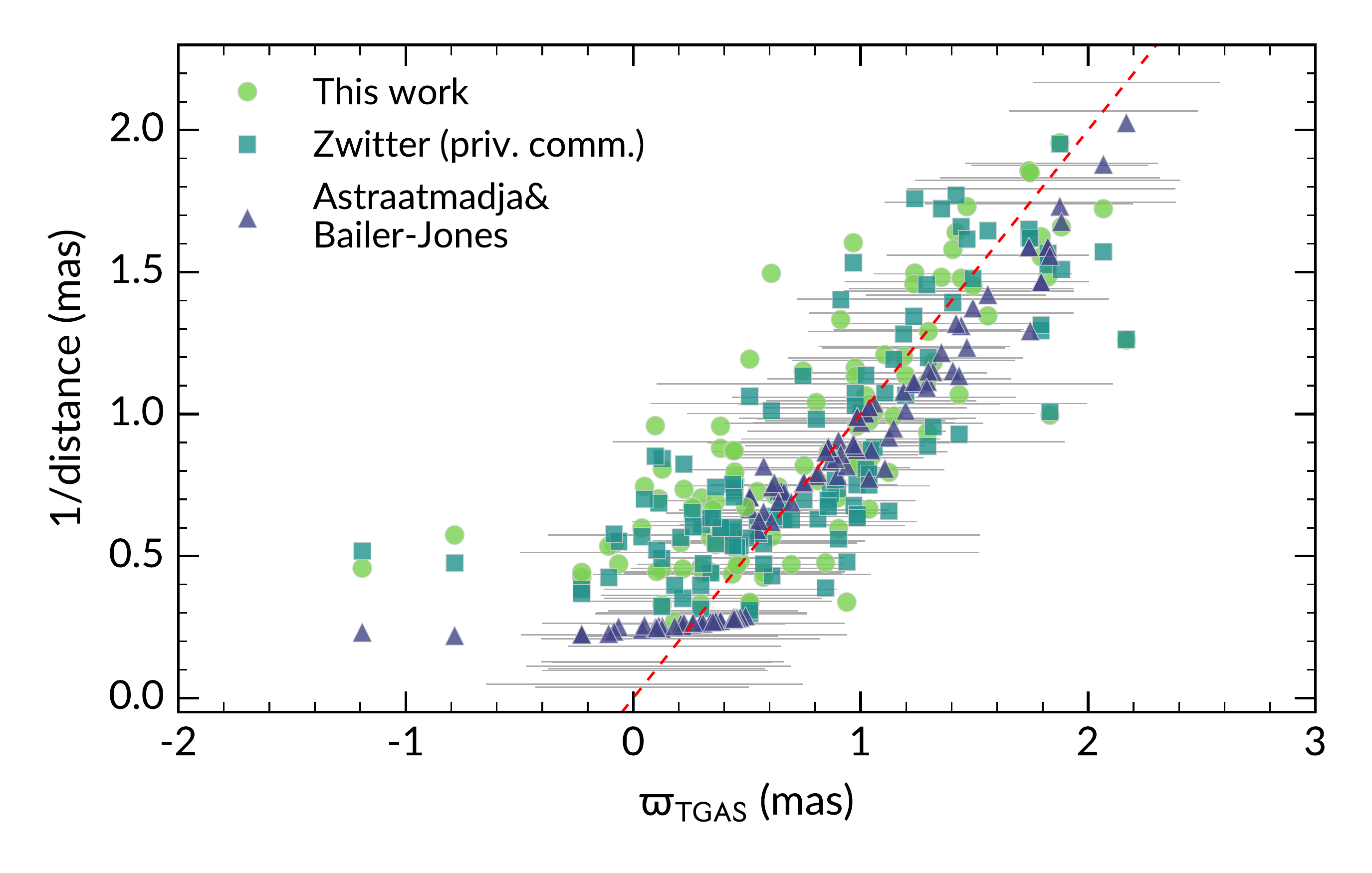}

	\caption{Comparison of distances derived from GALAH stellar parameters (this work/Zwitter) and Bayesian TGAS distances (Astraatmadja \& Bailer-Jones). On the $x$-axis are TGAS parallaxes, and the grey horizontal bars indicate reported uncertainties (including systematic errors). There is a break in the \protect\cite{Astraatmadja2016} values because a different prior is used for $\varpi<$ 0.5 mas, which according to the authors is more accurate for large distances (see text for details). The $y$-axis shows inferred parallaxes from the three distance methods. The dashed line is the 1:1 correlation. As the parallax decreases, TGAS fractional errors become very large, and in some cases negative parallaxes are reported. Compared to TGAS, distances derived from GALAH stellar parameters have an overall scatter of $\approx 0.37$ mas, which is within the typical TGAS uncertainty of 0.3 mas. For $\varpi<$ 0.5 mas, there is a systematic offset between TGAS and IRFM/Zwitter distances of $\approx 0.3$ mas.}
	\label{fig:dist-comp}
\end{figure*}

We tested the accuracy and precision of our distance determination method by comparing our results to the first $Gaia$ data release~\citep[][TGAS]{Brown2016}, which provides accurate parallaxes ($\varpi$) for bright stars in the $Tycho$-2 catalogue~\citep{Michalik2015}. Because of the brighter magnitude limit of $Tycho$-2, we only have a small overlap of about $100$ stars for comparison. We also compare our distances to those of~\cite{Astraatmadja2016}, who computed Bayesian distances using TGAS parallaxes and Milky Way density models. Finally, we include a comparison between our method and that of~\cite{Zwitter2010}, which computes the distance modulus by fitting stellar parameters to their most likely isochrone counterparts. 

Fig.~\ref{fig:dist-comp} compares the unaltered TGAS parallaxes with the inferred parallaxes from the three distance methods. Distances from~\cite{Astraatmadja2016} are median values of the posterior from their Milky Way density model. However, they note that the Milky Way model under-estimates distances for $\varpi<0.5$ mas when compared to Cepheid distances, as the model assumes that a star is more likely to be in the disk and photometric information is not used. Thus, the distances used here for $\varpi<0.5$ are the median of the posterior from their exponentially decreasing density model with scale length $L=1.35$ kpc~\citep{Astraatmadja2016}.

The comparison shows no systematic discrepancy for $\varpi_{\textrm{TGAS}}>0.5$ mas. The distances computed from our IRFM and the \cite{Zwitter2010} isochrone fitting method using the same spectroscopic parameters agree to within $\approx$$15$\%. Compared to TGAS, both the IRFM and isochrone fitting method have a standard deviation of $0.3$ mas, which is well within the typical errors quoted for TGAS parallaxes. We noticed that the Bayesian distances from~\cite{Astraatmadja2016} are slightly over-estimated compared to TGAS between $\varpi_{\textrm{TGAS}}$ = 1--2 mas. 

We do find an offset between all distance methods and the TGAS parallaxes for $\varpi_{\textrm{TGAS}}$ = 0--0.5 mas, where the TGAS values may be underestimated. For stars with $\varpi_{\textrm{TGAS}}>0$, the offset is 0.33 mas for the spectroscopic distances. \cite{Stassun2016} found a similar offset, between TGAS and inferred parallaxes derived from eclipsing binaries, however their results are applicable only to smaller distances, which is not seen in our results (see also \citealt{Huber2017}). A likely reason for the GALAH-TGAS offset is that TGAS uncertainties becomes very large at $\varpi < 0.5$ mas, so for a magnitude-limited sample, TGAS systematically scatters to smaller values. 

In summary, we find that our distances are accurate compared to TGAS parallaxes and the Bayesian distances of~\cite{Astraatmadja2016}, albeit with an offset for $\varpi_{\textrm{TGAS}}<0.5$ mas. Overall the standard deviation between the two spectroscopic methods is $17$\%. {Since both our method and the isochrone fitting method used the same set of stellar parameters, the comparison between them is indicative of their intrinsic uncertainties. Assuming that both methods contribute equally to the overall scatter, the internal uncertainty of each method is 12\%. This is the value we adopted as the our distance errors.}

\subsection{Separating the thin and thick disk populations}
\label{gmm}
High resolution spectroscopic studies show that the $\alpha$-enhancement of {local} disk stars follow two distinct tracks~(e.g., \citealt{Bensby2014,Adibekyan2011}). The high-$\alpha$ population is typically associated with the thick disk and has high velocity dispersion; the low-$\alpha$ stars are associated with the thin disk, with low velocity dispersion. The thick disk also has a larger rotational lag compared to the thin disk.

The thick disk can also be defined geometrically by star counts~\citep{Juric2008,Chen2011}, or by metallicity and kinematics~\citep{Katz2011,Kordopatis2011}. The thin and thick disks do overlap in their spatial, metallicity and kinematical distributions. Because of the two distinct sequences in [$\alpha$/M]-[Fe/H] space, definition of the thick disk by its enhanced \afe-abundances relative to thin disk stars of the same metallicity is currently widely used~\citep{Adibekyan2013, Bensby2014, Haywood2015} . However, the adopted dividing line between the high and low-\afe~populations differs from author to author. Furthermore, some stars with thick disk chemistry have thin disk kinematics, and there are stars that lie in the intermediate region between the two [\afe/M] sequences. The `thick disk' population that we are interested in is the stellar fossil of the turbulent epoch of fast star formation at high-$z$. Following this definition, we want to exclude the flaring outer thin disk, which contributes to the geometrical thick disk, and metal-rich stars which may have migrated from the inner thin disk.

To this end, we chose to separate the two components by fitting a mixture of Gaussian distributions using the Expectation-Minimisation algorithm~\citep{Dempster77}. We use three variables: [M/H] and [\afe/M] and the radial velocity (RV), { which, to the best of our knowledge, has not been done previously}. At $\ell=270^{\circ}$, the component of the rotational lag between the thin and thick disk along the line-of-sight is maximised for most of our fields, such that RV is a good proxy for $V$ velocity (see also \citealt{Kordopatis2017}). Instead of using the Cartesian $V$ space velocity component, which has significant proper motion errors, we use the precise GALAH radial velocity to help separate the two populations (98\% of our survey stars have RV uncertainty $<0.6$ kms$^{-1}$ according to~\citealt{Martell2017}). 

The \textsc{python} \verb|scikit-learn|~\citep{scikit-learn} module \verb|GaussianMixture| was used to perform the fitting. we assume that the data cube can be described by 2 multivariate Gaussians, each characterised by its three means and $3 \times 3$ covariance matrix:
$\theta_j = (\mu_j,\Sigma_j)$, where $j = \{1,2\}$, to represent the low and high-\afe \ sequences. Note that~\cite{Rojas-Arriagada2016} argue the [\afe/M] vs [M/H] distribution could be described by five components, but here we are not concerned with finding sub-components of the two \afe-sequences. 

Given a set of data ($x_1, x_2,..., x_n$), the likelihood function is defined as:
\begin{equation}
\mathcal{L}(\theta;x) = \prod_{i=1}^n \sum_{j=1}^2 w_jf(x_i;\mu_j,\Sigma_j),
\end{equation}
\noindent where $f$ is the probability density function of a multi-variate normal distribution and $w_j$ is the weight of each distribution. The algorithm initialises with random guesses for $\theta=(w_j,\mu_j,\Sigma_j)$ and iterates until the log-likelihood is at minimum. The probability that a data point $x_i$ belongs to component $j$ is given by:
\begin{equation}
P_j\,(x_i\,|\,\theta)=\frac{w_jf(x_i;\mu_j,\Sigma_j)}{w_1f(x_i;\mu_1,\Sigma_1)+w_2f(x_i,\mu_2,\Sigma_2)}
\end{equation}
\noindent where
\begin{equation}
P_1(x_i) + P_2(x_i) = 1
\end{equation}

\begin{figure*}
\centering
		\includegraphics[width=1.2\columnwidth]{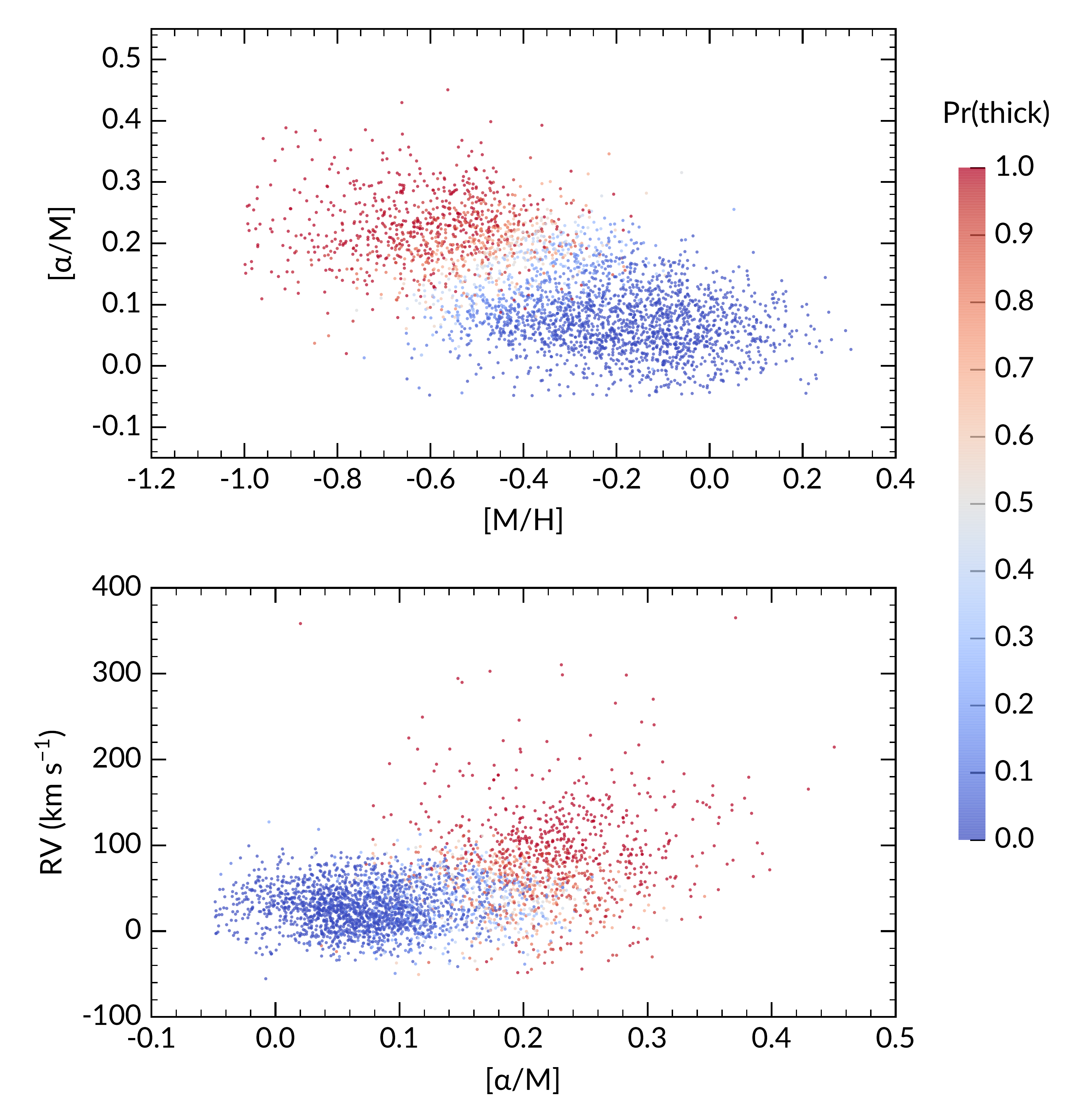}

	\caption{Results of the Gaussian mixture decomposition. Top panel: projection along the [\afe/M]--[M/H] plane. Bottom panel: projection along the [\afe/M]--RV plane, both colour-coded by the probability of a star belong to the thick disk. We can see that there are two well defined populations in both projections, however there are also plenty of stars that are difficult to place in either population. These stars have chemistry and kinematics that could belong to either of the classically defined `thin' and `thick' disk. Stars that are typically defined as thin disk by chemistry have Pr(thick) $\leq$ 0.1. Stars with typical thick disk chemistry however have a higher velocity dispersion and therefore a larger spread in probability, ranging from 0.6 $\leq$ Pr(thick) $< 1$.}
	\label{fig:gmm}
\end{figure*}

Fig.~\ref{fig:gmm} shows projections in the [$\alpha$/M]--[M/H] and the RV--[$\alpha$/M] planes, where two Gaussian components centred at [$\alpha$/M] = 0.05 and [$\alpha$/M] = 0.2 can be seen, each with a distinctive median radial velocity. Stars are colour-coded by their thick disk probability. As expected, the high-\afe~stars have much higher thick disk probability than the low-\afe~stars. It is also apparent that stars with [M/H] between $-0.4$ and 0 are likely to be designated thin disk membership, because their radial velocities and metallicity more closely resemble that of thin disk stars. This is perhaps the most important distinction between our `thick disk' definition and that of other studies: that the overall more \afe-enhanced population does not include the high metallicity, high-\afe~population. This places important constraints on subsequent analyses and the interpretation of our results with respect to models of disk formation, as here we are assuming that the thick disk is almost exclusively old by excluding more metal-rich stars. 

Furthermore, Fig \ref{fig:gmm} indicates that there are `transition stars', which have higher [\afe/M] than thin disk stars at the same metallicity and kinematics that lie between the two disks, making it is difficult to assign them to either population. We therefore assigned thin disk membership only to stars that have thick disk probability $\leq 0.1$, which have [\afe/M] 
$\leq 0.15$, consistent with the location of the `gap' between high and low \afe~populations for our data set. The majority of stars with thick disk probability between 0.1--0.5 have [\afe/M] values between 0.15--0.3 dex. These `transition' stars are omitted from the analysis to minimise contamination in each defined population. Approximately 13\% of the overall sample are in the `transition' category. 

\section{Selection bias} 
\label{bias}

Fig. \ref{fig:field_loc}, right panel, shows that the pilot survey has a simple magnitude cut, 10 $<K_s$ $<$ 12, while the main survey colour-magnitude selection appears to be a stripe in the $(J-K_s)_0$ vs $K_s$ plane, from the criterion that  $12 < V_{JK}< 14$. The main survey also observed some bright stars that fall outside the lower magnitude limit. In addition, the pilot survey purposely observed a larger fraction of stars at higher latitudes, which means that the population at low latitudes is under-represented. These selection biases affect the resulting metallicity, distance (and therefore vertical height) distributions of the observed population. In this section we aim to correct for these effects so that the underlying Galactic population can be correctly recovered.

\subsection{Correcting for selection effects}
\subsubsection{Field selection bias}
The first selection effect that we corrected was the bias from targeting particular fields. We purposely observed a larger relative fraction of stars at higher latitudes to target the thick disk, and thus biased against low latitude stars. 

To correct for this, we determined for each field the number of stars present in the observed sample compared with the number of photometric targets available for that field in the GALAH input catalogue, within the same magnitude limit, e.g.: 

\begin{equation*}
w_{field}=\frac{N_{observed} \ (12<{V}_{JK}<14)}{N_{targets} \ \ (12<{V}_{JK}<14) } 
\end{equation*}

We dealt with the magnitude ranges of the pilot and main surveys separately. The limits used are 12 $<$ V$_{{JK}}$ $<$ 14 for main survey fields; 12 $<$ V$_{{JK}}<$14.5 for pilot fields and 9 $<$ V$_{{JK}}$ $<$ 12 for the bright field.

\subsubsection{Magnitude and colour selection bias}

\begin{figure*}
	\centering
	\includegraphics[width=0.6\textwidth,trim={4.8cm 1cm 0.8cm 1cm},clip]{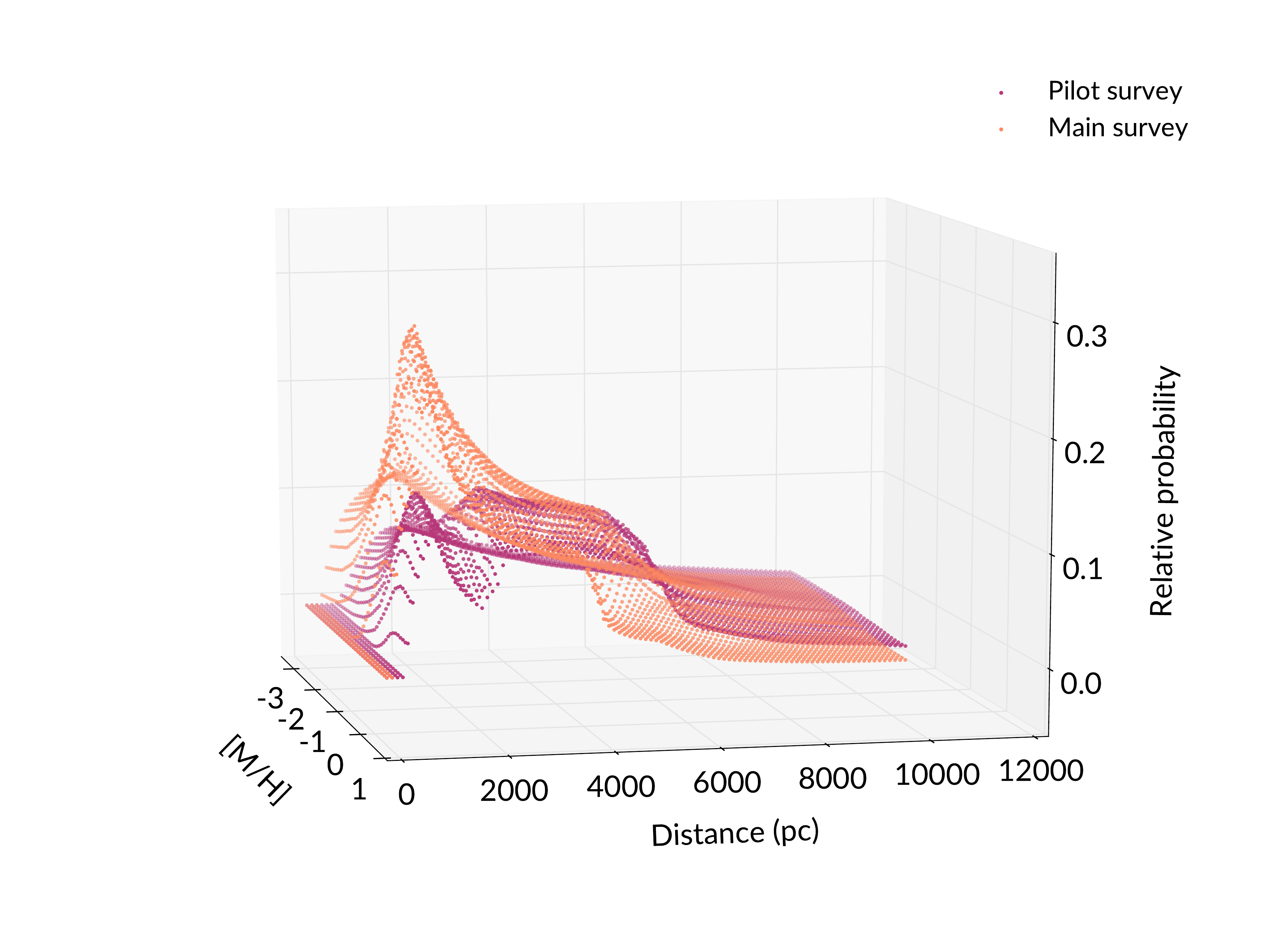} 
	
	\caption{The relative probability of observing a particular star in the metallicity-distance plane given the colour-magnitude selection of GALAH pilot and main surveys. While we show the metallicity distribution up to [M/H]=$-$3 dex, we do not have any stars with [M/H] $<$ $-$1 in our sample. The distance distribution is most affected by the colour-magnitude selection of the two surveys, with the pilot survey favouring more distant stars. The metallicity of both surveys peaks around solar, but compared to the main survey, the pilot survey has a larger fraction of stars that are more metal-poor.}
	\label{fig:isopr}
\end{figure*}

Following~\cite{Casagrande2016}, we assessed the magnitude and colour selection bias by creating a synthetic population using BaSTI isochrones~\citep{Pietrinferni2004}. From a data cube that spans 0.5 to 10 Gyr in age, $-3$ to 0.5 dex in metallicity and 10 to 10000 pc in distance, each point in the age and metallicity plane is populated on the isochrones according to the \cite{Salpeter1955} IMF, with the distances providing apparent magnitudes for each population. We then applied the same apparent colour and magnitude cut as shown in Fig. \ref{fig:field_loc} to obtain the ratio of stars observed with our selection function compared to the total number of stars that populate a given point in the age, metallicity and distance cube. As in the previous section, the pilot and main survey selection functions are taken into account separately. Because there is no age information available for this sample, we integrated the observed probabilities over all ages for each point in the distance--metallicity plane. This implicitly assumes that the age distribution is flat in the solar neighbourhood (e.g., {\citealt{Edvardsson1993,Ting2015a}}). With this method, the effects of different evolutionary time-scales of each stellar population on the HR diagram are also taken into account via the IMF. 

Fig. \ref{fig:isopr} shows the relative fraction of stars observed after the colour-magnitude selection is applied. The most metal-poor and metal-rich stars are slightly biased against, similarly so for both the pilot and main survey selections. The distances, on the other hand, are very different for the pilot and main surveys. The main survey is biased against stars more distant than 1.5 kpc, especially at lower metallicities. The pilot survey observes relatively more distant (and thus larger $|z|$) stars as intended. In addition, the pilot survey colour and magnitude limit particularly targetted red clump stars, which primarily contributed to the second peak in its selection function. 

The relative ratios obtained from this population synthesis method are dependent on the choice of stellar models and IMF, however we note that we are only using these numbers in the relative sense, to gauge the importance of one star compared to another. In this sense, we do not expect the selection effects to change qualitatively. 

\subsection{Effects of bias correction}
The final weight is determined by combining the fraction from field selection bias and the isochrone population synthesis. Since the fraction indicates how likely a star is observed, the weights are computed as
\begin{equation*} w_{final} = \frac{1}{w_{field} \times w_{isochrone}}
\end{equation*}
so that stars less likely to be observed are given higher weights. 

\begin{figure*}
	\centering
	\includegraphics[width=0.5\textwidth,trim={0.5cm 0.9cm 0.9cm 0.9cm},clip]{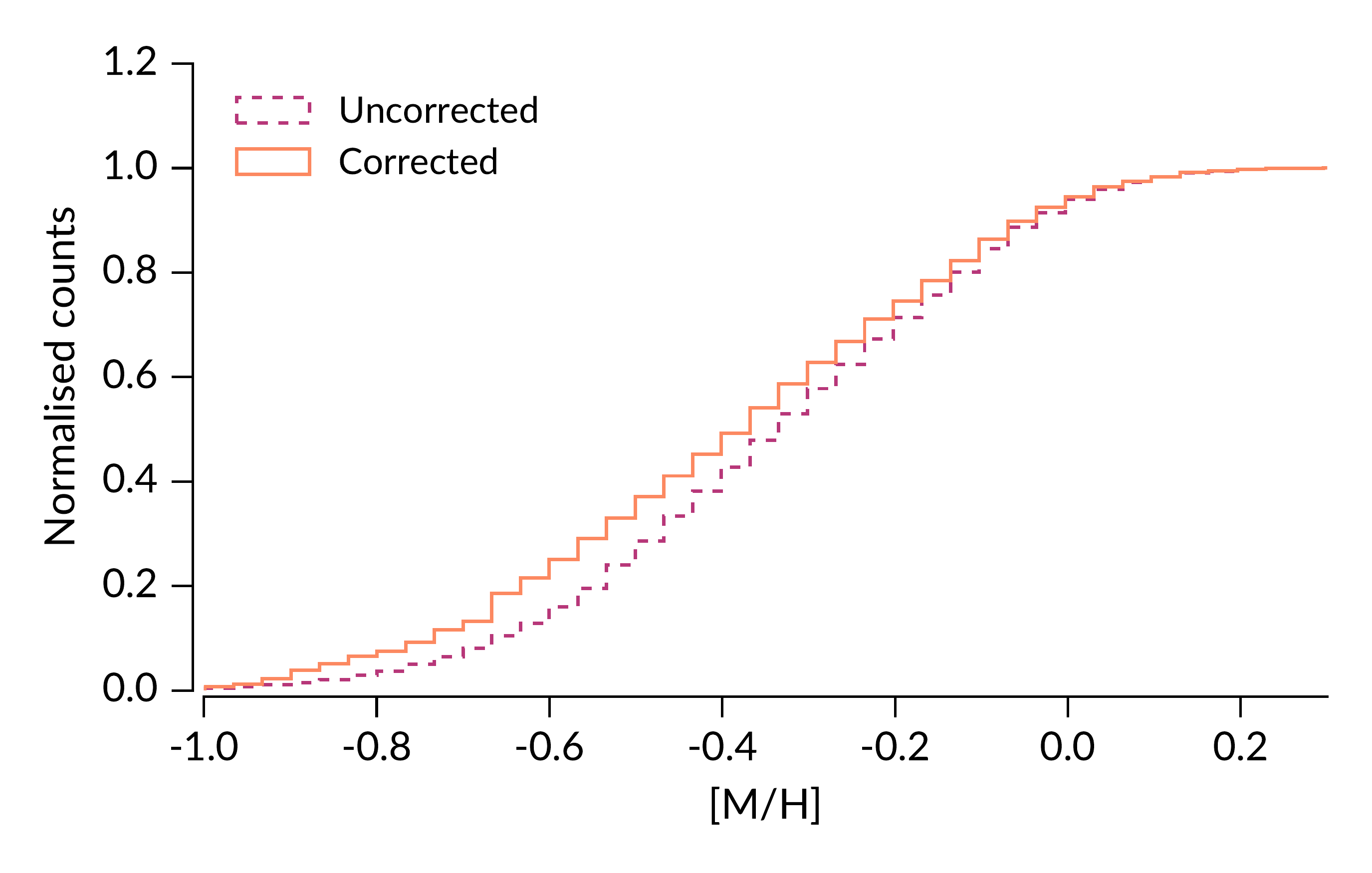}\includegraphics[width=0.5\textwidth,trim={0.5cm 0.9cm 0.9cm 0.9cm},clip]{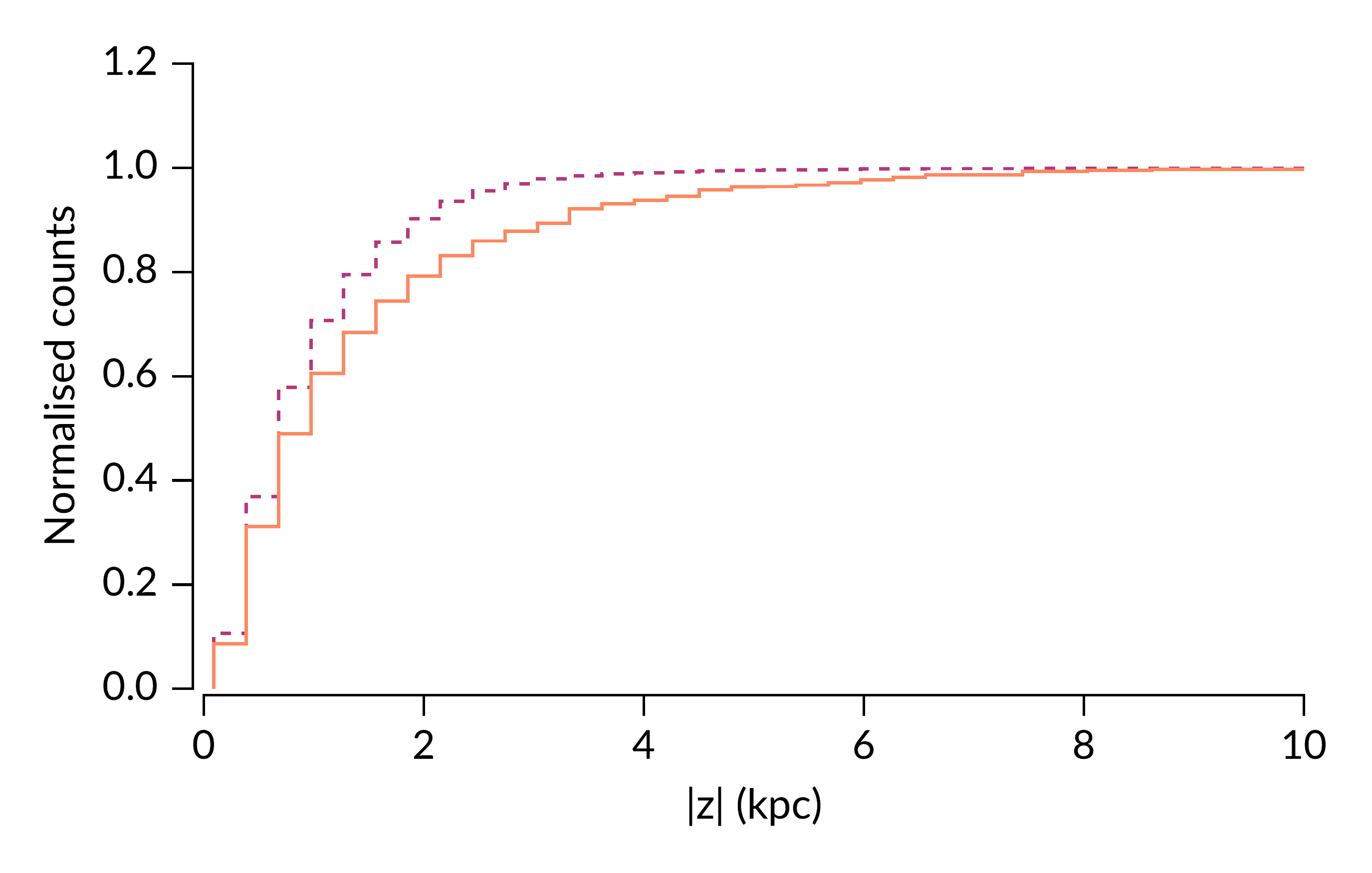}
	\caption{Cumulative histograms showing bias-uncorrected and corrected metallicity and height $|z|$ distributions. Left panel: [M/H] distribution. Right panel: height distribution. Typically the corrections account for the bias against stars that are more metal-poor and further from the plane.}
	\label{weighted}
\end{figure*}

Overall, the corrections mean that more metal-poor and distant stars are weighted more heavily. Effects of the weights on the [M/H] and $|z|$ distributions are shown in Fig. \ref{weighted}.

\subsection{Halo contamination}

To assess the halo contamination in our low and high-\afe~samples, we used the \textit{Galaxia} code~\citep{Sharma2011}, based on the \textit{Besan\c{c}on} models~\citep{Robin2003} to synthesise the stellar population within our observed region. We applied the same colour-magnitude limits (in 2MASS $J,K_s$ photometry as shown in Fig. \ref{fig:field_loc}) for the pilot and main survey samples separately. The simulation shows that within our metallicity range $(-1 \leq$ [M/H] $\leq 0.4)$, the contamination of halo stars is extremely small, at 0.5\% for both of the pilot and main surveys. Therefore, any effects of halo stars on our results would be negligible. 

\section{Metallicity profiles} 
\label{gradients}
	
\subsection{Radial metallicity profiles}

\begin{figure*}
	\includegraphics[width=0.5\textwidth,trim={0.6cm 0.9cm 0.9cm 0.9cm},clip]{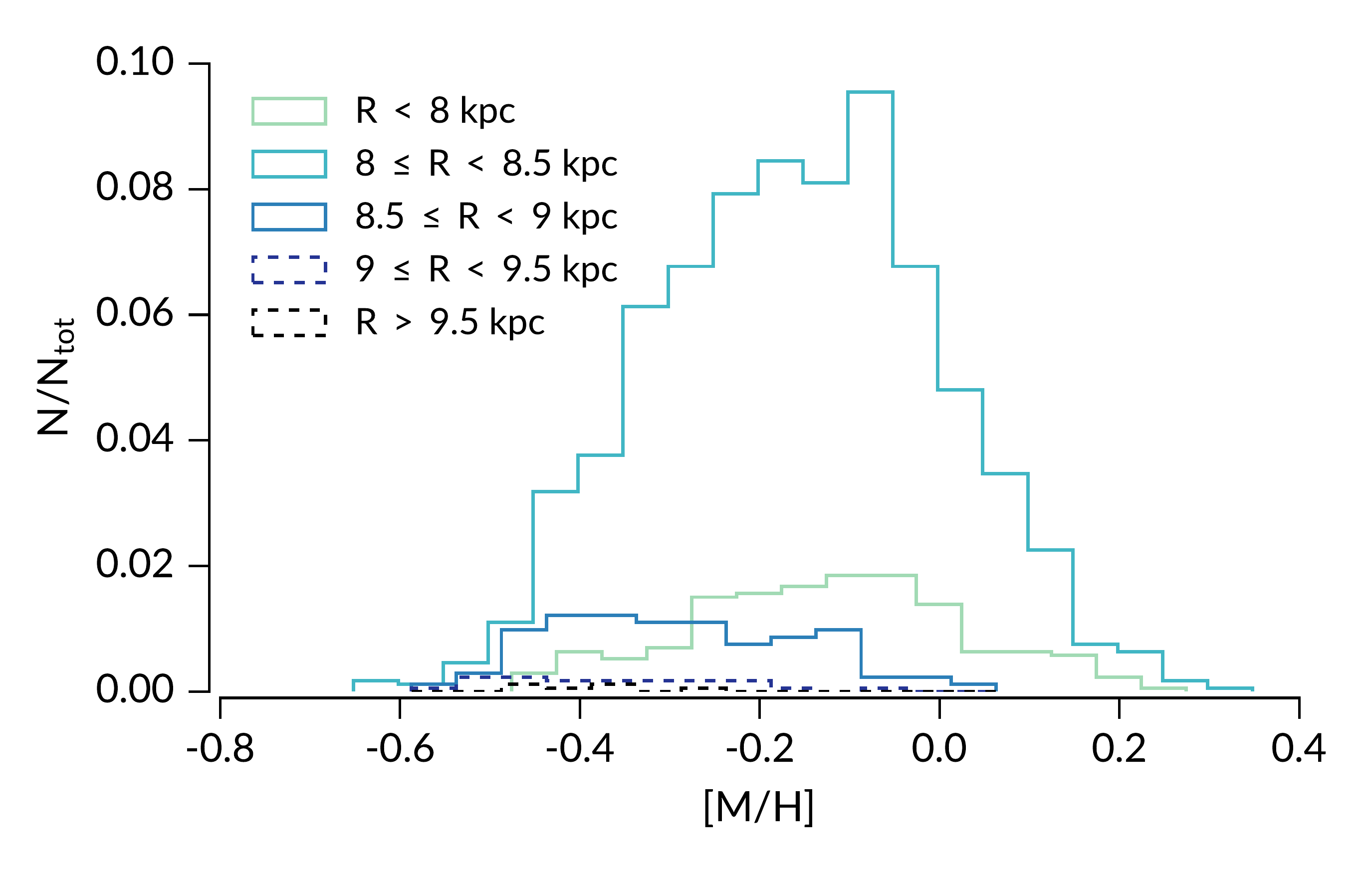}\includegraphics[width=0.5\textwidth,trim={0.6cm 0.9cm 0.9cm 0.9cm},clip]{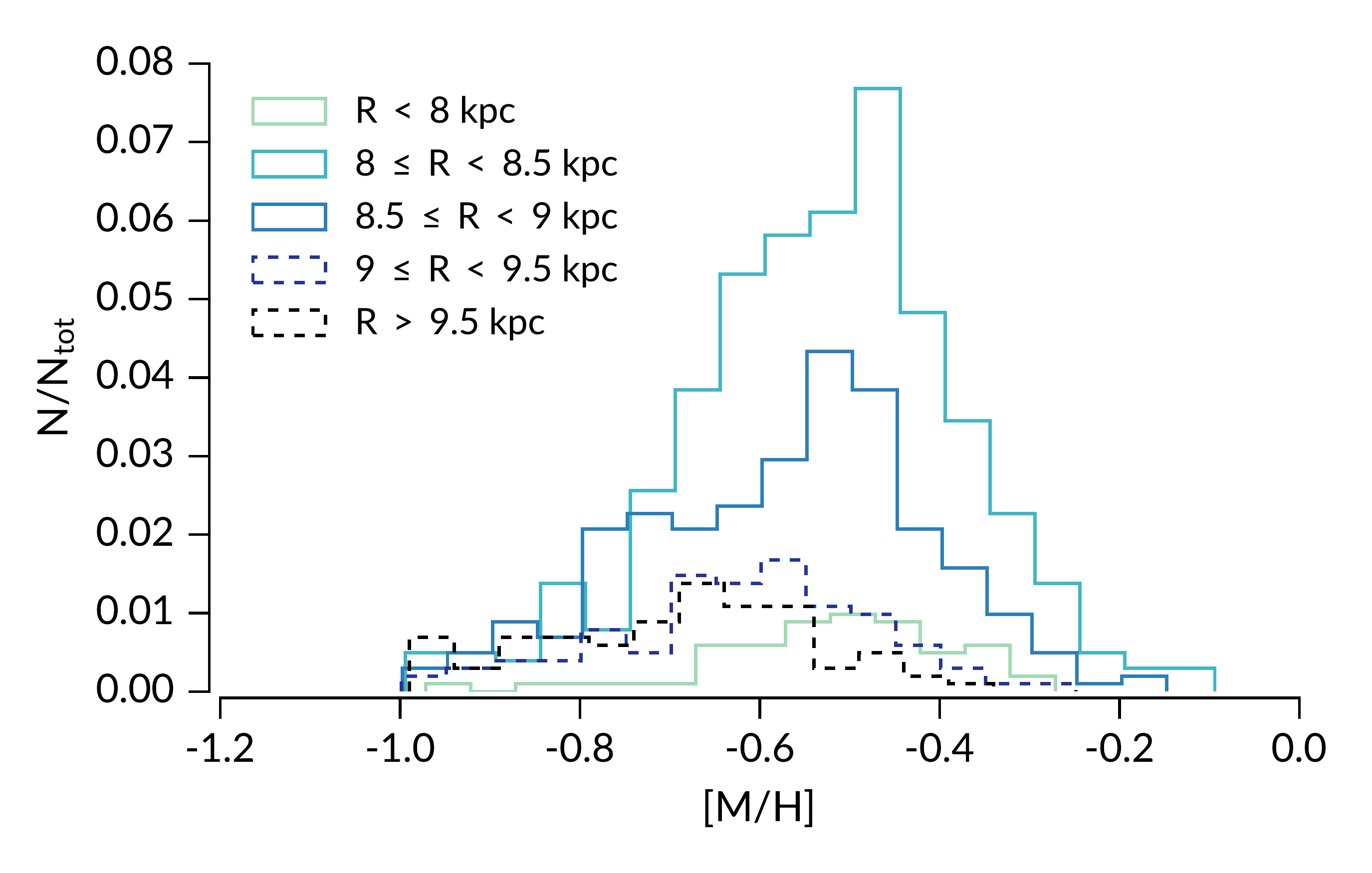}
	\caption{Radial distribution of the thin (low-\afe) and thick (high-\afe) disks, {after correcting for selection effects}. Left panel: the thin disk's mean metallicity changes rapidly as a function of radial distance. This is due to both the radial metallicity gradient observed in the thin disk, and that the average vertical height increases with increasing radial distance. Right panel: The thick disk, on the other hand, does not show a strong change in shape nor median value with radial distance.}
	\label{fig:abundradthin}
\end{figure*}

In figure~\ref{fig:abundradthin} we show the MDF of the thin and thick disks in radial distance bins of 500 pc. Within the small range that we cover, no metallicity gradient is observed for the thick disk. The MDF remains constant in shape and median value across all radial distances up to 9 kpc, which is consistent with~\cite{Hayden2015}. Beyond 9 kpc, we notice that the MDF skews slightly towards more metal-poor values, but we interpret this as an effect caused by observing progressively larger median $|z|$ as we move further from the Galactic centre (see Fig. \ref{fig:Rz}) rather than the thick disk having a radial metallicity gradient. 

For the low-\afe~population, we observe only a small number of stars at radial distances further than 9 kpc. The distribution is roughly Gaussian at all locations, but skews towards metal-poor with increasing $R$. In Fig.~\ref{fig:Rz}, it is evident that the majority of \afe-poor stars (thin disk) are confined to the plane. At $R_\mathrm{GC}=8.5$ kpc, most of the stars lie above |$z$| = 1 kpc, so here we are likely to be observing only the metal-poor tail of the thin disk. The shift towards lower metallicity at large $R$ is likely due to the radial metallicity gradient of the thin disk~{\citep{Cheng2012, Genovali2014, Hayden2015}}, and the vertical gradient discussed in the next section.

\begin{table}
	\caption{The mean (and standard error), dispersion and skewness of the metallicity distributions at different radial bins for the low-\afe~population.}
	\label{table:normdist}
	\begin{tabular}{llll}
		\hline 
		Radial range & Mean [M/H] & Standard deviation & Skewness \\
		\hline
		R $<$ 8 kpc & $-0.12 \pm 0.01$ & 0.15 & $+0.01 \pm 0.02$  \\
		8 $\leq$ R $<$ 8.5 kpc & $-0.16 \pm 0.004$ & 0.16 & $-0.02 \pm 0.01$ \\
		8.5 $\leq$ R $<$ 9 kpc & $-0.30 \pm 0.01$ & 0.13 & $+0.30 \pm 0.02$ \\
		9 $\leq$ R $<$ 9.5 kpc & $-0.37 \pm 0.03$ & 0.17 & $+1.50 \pm 0.07$ \\
		R $>$ 9.5 kpc & $-0.42 \pm 0.05$ & 0.13 & $-0.30 \pm 0.07$\\
		\hline
	\end{tabular}
\end{table}

\begin{table}
	
	\caption{The mean (and standard error), dispersion and skewness of the metallicity distributions at different radial bins for the high-\afe~population.}
	\label{table:normdist2}
	\begin{tabular}{llll}
		\hline 
		Radial range & Mean [M/H] & Standard deviation & Skewness \\
		\hline
		R $<$ 8 kpc & $-0.52 \pm 0.02$ & 0.13 & $-0.79 \pm 0.04$\\
		8 $\leq$ R $<$ 8.5 kpc & $-0.53 \pm 0.01$ & 0.15 & $-0.27 \pm 0.01$ \\
		8.5 $\leq$ R $<$ 9 kpc & $-0.57 \pm 0.01$ & 0.15 & $-0.26 \pm 0.02$ \\
		9 $\leq$ R $<$ 9.5 kpc & $-0.61\pm 0.01$ & 0.14 & $-0.36 \pm 0.03$ \\
		R $>$ 9.5 kpc & $-0.70\pm 0.02$ & 0.15 & $-0.07 \pm 0.04$\\
		\hline
	\end{tabular}
\end{table}

{We provide the median, standard deviation and skewness of each radial bin of each population in tables \ref{table:normdist} and \ref{table:normdist2}. These support our conclusions that the MDFs of both populations within our $R_\mathrm{GC}$ range are close to Gaussian, and that there is little change in the mean metallicity and shape of the high-\afe~population. We caution that the statistics is more uncertain for bins $R_\mathrm{GC} > 9$ kpc of the low-\afe~population due to the small sample size.}

\subsection{Vertical metallicity profiles}

\begin{figure}
\centering
		\includegraphics[width=1.05\columnwidth,trim={0.7cm 1cm 0cm 0cm},clip]{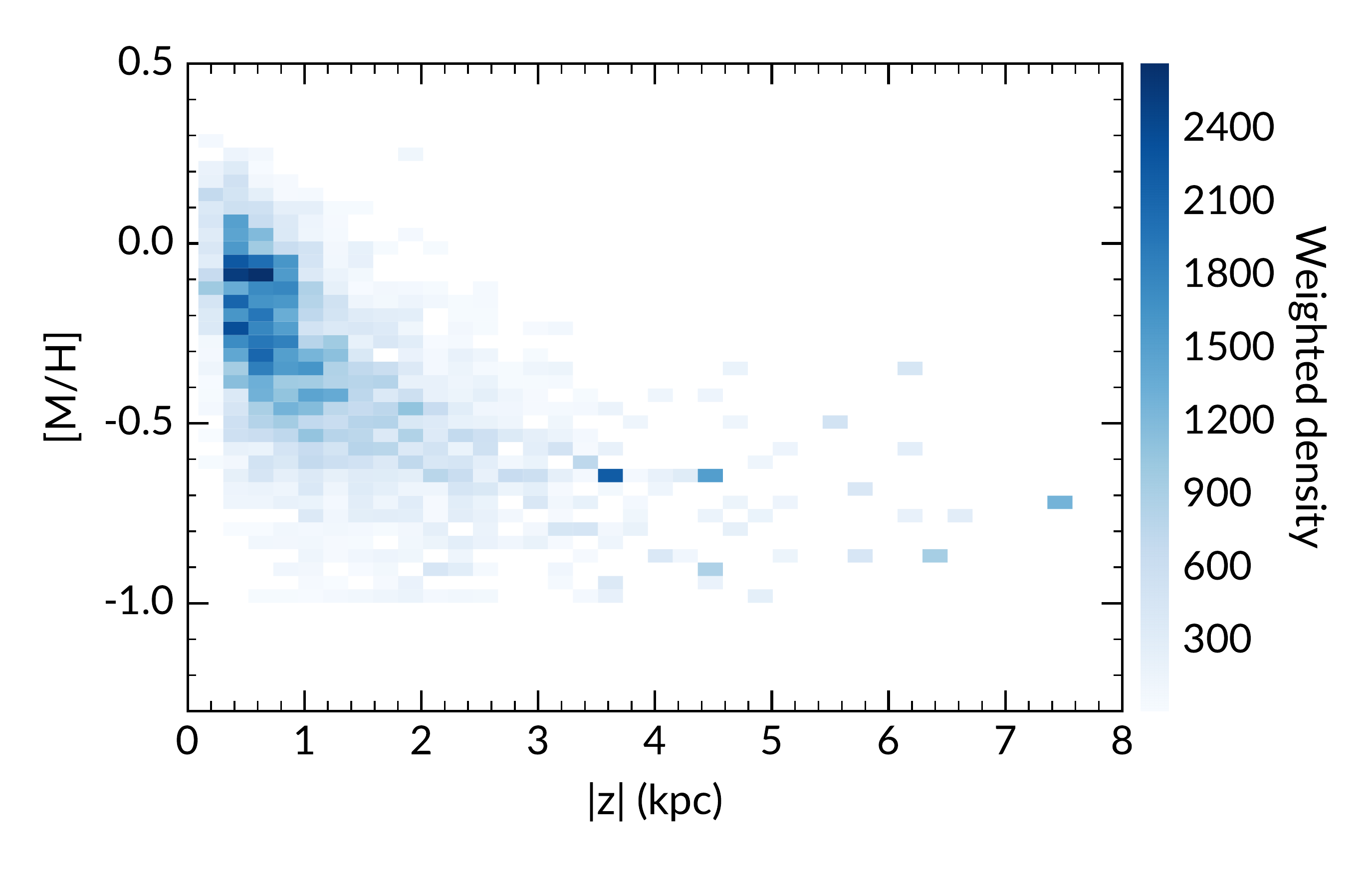}

	\caption{Variation of metallicity with distance from the Galactic plane for all stars independently of their alpha assignment (the `transition' stars mentioned in Section \ref{gmm} are also included). The density was weighted using bias correction fractions described in Section \ref{bias}. The metallicity decreases smoothly with increasing height, however the gradient appears to flatten at |\textit{z}| = 2 kpc.}
	\label{fig:fehgrad_all}
\end{figure}

The vertical gradients were measured using an orthogonal linear least squares regression to all data points, taking into account each data point's uncertainties in [M/H] and vertical height. Each point is then weighted by the selection bias correction described in Section \ref{bias}. We do this by decreasing the uncertainty of each data point by the square root of the correction factor. In this section we report the gradients measured for the disk as a whole, and for each defined \afe~sub-population. The gradients measured are summarised in Table \ref{table:1}.

Fig. \ref{fig:fehgrad_all} shows a density plot of the metallicity as a function of height above the plane for all stars, including those that were omitted from the individual \afe-subpopulation, as explained in Section \ref{gmm}. The density was weighted to correct for selection biases using relative fractions described in Section \ref{bias}. We observe that the metallicity decreases smoothly as $|z|$ increases. The vertical gradient for the disk overall is d[M/H]/d$z = -0.22 \pm 0.01$ \gradunit, and appears to flatten at larger $|z|$, from about $|z| = 2$ kpc. The gradient value is in good agreement with overall disk gradients measured by \cite{Schlesinger2014} for a sample of volume-complete SEGUE dwarfs. 

\begin{figure}
\centering
	\includegraphics[width=1.04\columnwidth,trim={0.8cm 1cm 0cm 0cm},clip]{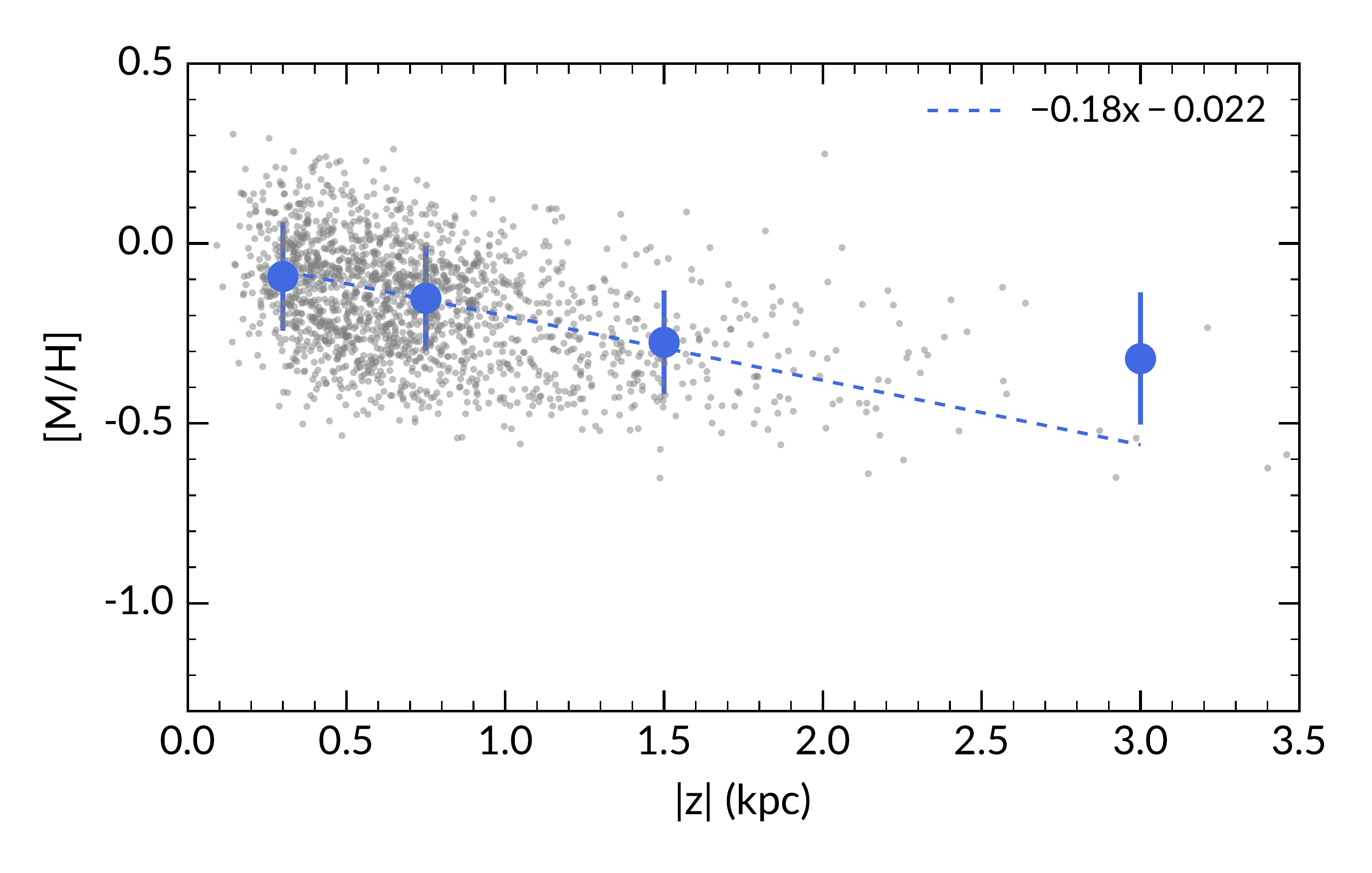}\\
	\includegraphics[width=1.03\columnwidth,trim={0.8cm 1cm 0cm 0cm},clip]{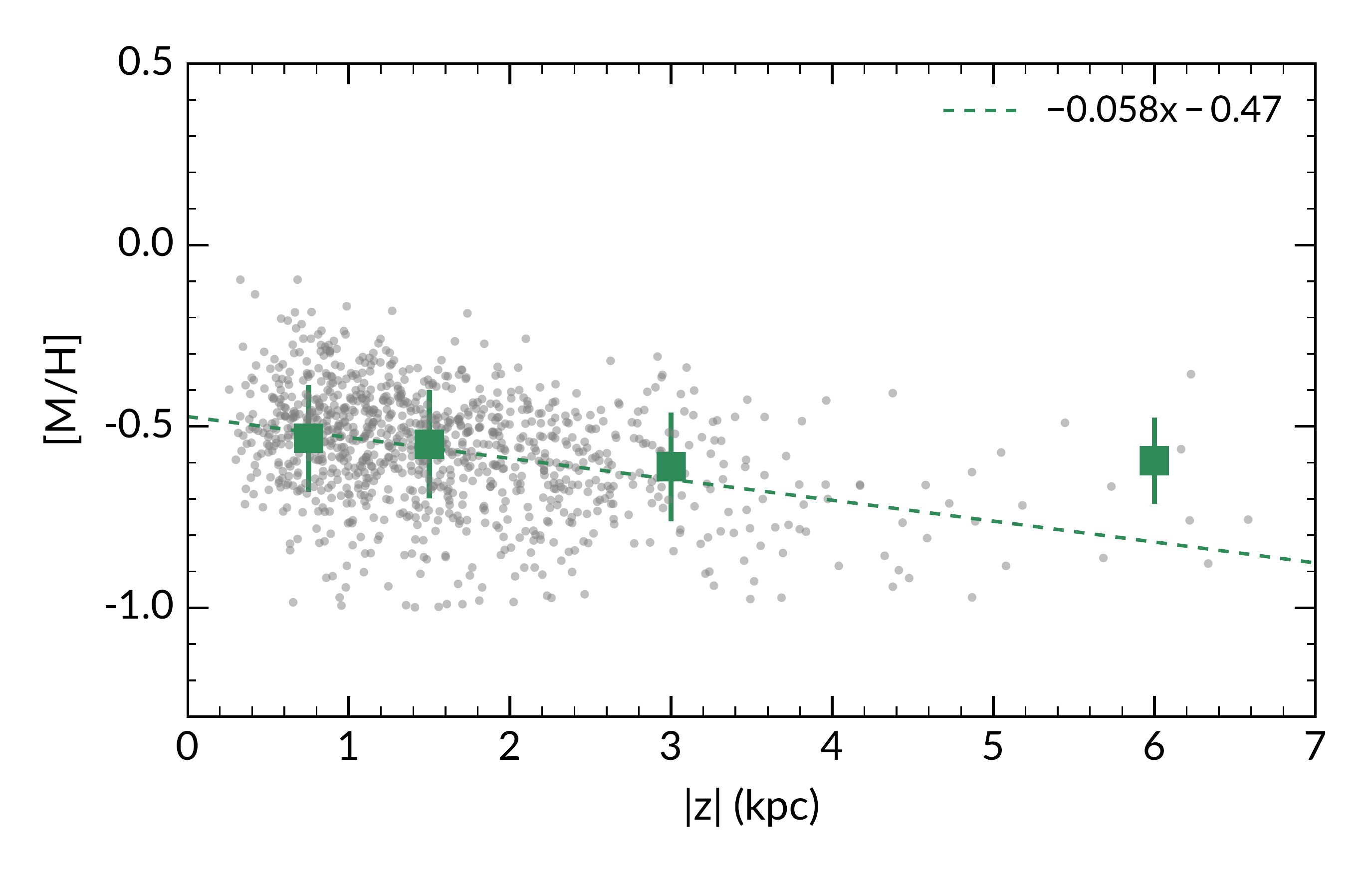}

	\caption{Variation of metallicity with distance from the Galactic plane for each \afe~sub-population. Top panel: the thin disk has a steep negative gradient, which is consistent with what many authors have observed previously. Bottom panel: the thick disk has a shallower gradient. The trends are fitted over grey data points, over-plotted are averaged values of four height bins and their one sigma error bars. It is important to note that the binned values were not used in the gradient fitting.}
	\label{fig:abundzthin}
\end{figure}

For each of the sub-populations, we also found a metallicity gradient, as shown in Fig.~\ref{fig:abundzthin}. Over-plotted in each panel are averaged values of metallicity at different $|z|$ bins for clarity, but these binned values have no effect on the data fitting. We discuss the vertical gradients below.

\subsubsection{The low-\afe~population}

The low-\afe~population, or thin disk, is known to have a radial metallicity gradient d[M/H]/d$R$ of $\approx$$-0.08$ \gradunit, which flattens at progressively higher $|z|$~\citep{Cheng2012,Hayden2014}. The radial metallicity gradient can be seen in the left panel of Fig. \ref{fig:abundradthin}, where the median metallicity shifts to lower values at larger $R_\textrm{GC}$. The small $R_\textrm{GC}$ range that we cover does not allow us to reliably measure radial metallicity gradients, so we corrected for this effect by estimating the metallicity of each star at $R=8$ kpc using radial gradients specified in~\cite{Cheng2012} for height bins $0.25< |z| < 0.5$ kpc; $0.5<|z|< 1$ kpc and $1<|z|< 1.5$ kpc. The data set of~\cite{Cheng2012} did not extend beyond  $|z|$ = 1.5 kpc, so for all heights above this value, we assumed the same radial gradient as at $1<|z|<1.5$. Overall, the radial gradient correction caused a change of $-0.01$ \gradunit in the vertical gradient. The final weighted vertical gradient of the low-\afe~population d[M/H]/d$z$ = $-0.18 \pm 0.01$ \gradunit.

Studies that were conducted prior to recent large spectroscopic surveys typically reported steeper negative gradients than our value. \cite{Bartasiute2003} separated thin disk stars by rotational lag and measured $-0.23 \pm 0.04$ \gradunit. \cite{Marsakov2006} used both space velocities and orbital eccentricities restrictions to select thin disk stars and reported a gradient of $-0.29 \pm 0.06$ \gradunit. It is highly likely that separating the thin disk purely based on kinematics would result in contamination of thick disk stars, which explains why these gradients are in agreement with our overall disk gradient, but steeper than the gradient of the low-\afe~population. 

Few {studies of disk vertical metallicity gradients} separated the thin/thick disk using chemistry. The only recent studies that identified the thin disk by their \afe-abundances are \cite{Schlesinger2014} (SEGUE), \cite{Hayden2014} (APOGEE) and \cite{Mikolaitis2014} (\textit{Gaia}-ESO).~\cite{Hayden2014} found a low-\afe~gradient of $-0.21 \pm 0.02$ \gradunit~at the solar circle for the APOGEE DR10 sample, which is slightly steeper than our value. \cite{Hayden2014} also correct for the radial metallicity gradient, using values similar to that of \cite{Cheng2012} used here. The small discrepancy could arise from our different definitions of the thin disk, as \cite{Hayden2014} made a straight-line cut at [\afe/M] = 0.18. In Fig. 6 of \cite{Hayden2014}, their low-\afe~population extends to [M/H] = $-2$ dex while ours extends to only [M/H] = $-0.6$ dex. The low-\afe, very metal-poor stars seen in APOGEE data could belong to the halo{~\citep{Nissen2010,Adibekyan2013}}, and this contamination would steepen the gradient. 

~\cite{Schlesinger2014} also computed a gradient for the low-\afe~population of SEGUE dwarfs. They measured, for the disk as a whole, a vertical metallicity gradient of $-0.24^{+0.04}_{-0.05}$ \gradunit, which is in agreement with our measurement. However their low-\afe~population has a gradient consistent with zero: d[M/H]/d$z=-0.01^{+0.09}_{-0.06}$ \gradunit. However, their intermediate \afe~sub-population with $0.2<$[\afe/M]$<0.3$ has d[M/H]/d$z=-0.17^{+0.08}_{-0.07}$ \gradunit, which agrees with our low-\afe~metallicity gradient. We thus conclude that the discrepancy between our result and that of \cite{Schlesinger2014} is largely due to the chemical separation criteria {(also see \citealt{Ciuca2018}, who found an age-dependence for the thin disk vertically metallicity gradient, such that the youngest population has a flatter gradient).}

~\cite{Mikolaitis2014} measured a slightly shallower gradient of d[M/H]/d$z=-0.11 \pm 0.01$ for \textit{Gaia}-ESO dwarfs and giants. The \textit{Gaia}-ESO sample is more metal-poor overall and \cite{Mikolaitis2014} separated thick disk stars by the location of under-densities in their [\ion{Mg}{I}/M] histograms (their Fig. 3). The dividing line is at different values of [\ion{Mg}{I}/M] for different metallicity regimes. In Fig. 10 of \cite{Mikolaitis2014}, it is clear that their sample is biased against metal-rich stars, such that there are very few stars with [M/H]$>0$ (also see~\citealt{Stonkute}).

\subsubsection{The high-\afe~population}
\label{subsubsec:highalphafeh}
The vertical metallicity distribution of the thick disk (high-\afe) stars, is relatively flat compared to the low-\afe~population, at $-0.058 \pm 0.003$~\gradunit. Several authors have measured the vertical gradient for the thick disk, using different methods to define this population. Earlier studies, such as \cite{Gilmore1995} and \cite{Prieto2006}, reported no vertical metallicity gradient in the thick disk (\citealt{Prieto2006} quoted an upper limit of d[M/H]/d$z=0.03$ \gradunit). {More recently, \cite{Boeche2014} concluded that the vertical metallicity gradient of the thick disk is consistent with zero, based on a sample of RAVE giants}. However, other studies, using a combination of metallicity or kinematics to separate the thick disk have reported a shallow metallicity gradient.

\cite{Katz2011} observed sub-giants at two lines of sight: $(l,b)=(51^{\circ},80^{\circ})$ and $(5^{\circ},46^{\circ})$ at low resolution. Their metallicity distribution functions show signs of bimodality, and the thick disk was defined as stars centred around [M/H] $\approx$ $-0.5$ dex. The vertical gradient measured by \cite{Katz2011} is $-0.068 \pm 0.009$ \gradunit, consistent with our value. 

\cite{Ruchti2011} observed a number of metal-poor thick disk candidates at high resolution using the MIKE, FEROS and UCLES spectrographs ($\lambda/\Delta \lambda \approx 35,000-45,000$). They classified their stars based on a Monte Carlo simulation of space motion $U, V, W$, assuming Gaussian errors on the velocities and distances. By further restricting their \afe-enhanced sample with thick disk kinematics to metal-poor stars only ([M/H]$\leq -1.2$), they avoid most thin disk contamination. The measured gradient is $-0.09 \pm 0.05$ \gradunit, which also agrees with our results.

\cite{Kordopatis2011} observed stars using the VLT/GIRAFFE spectrograph ($\lambda/\Delta \lambda \approx 6500$) at almost the same Galactic longitude as the GALAH pilot survey $(\ell=277^{\circ})$, and the same latitude as our highest fields 
$(b=47^{\circ})$. They reported a gradient of $-0.14 \pm 0.05$ \gradunit~for stars at heights $1<|z|<4$ kpc, where the thick disk is dominant, which does not agree with our result. Selecting the thick disk based only on height above the plane will certainly include thin disk contaminants and thus cause their gradient to be steeper. 

\cite{Chen2011} selected a sample of SDSS stars at $1<|z|<3$ kpc to represent the thick disk and measured a vertical gradient of $-0.22 \pm 0.07$ \gradunit. From the separation by chemistry shown in this paper and elsewhere, thin disk stars exist at |$z$| up to at least 2 kpc, so a thick disk definition based on vertical height alone is not very accurate.~\cite{Chen2011} provides another estimate of $-0.12 \pm 0.01$ \gradunit~for the gradient after they have modelled and subtracted thin disk contaminants using the Besan\c{c}on model, which is closer to our value. {However, neither of these thick disk vertical metallicity gradients is in agreement with our value.}

Comparing our measurement of the vertical gradient for the high-\afe~population with the gradient from the APOGEE DR10 \citep{Hayden2014} reveals a large discrepancy, as they found a steep negative gradient of $-0.26 \pm 0.02$ \gradunit~at the solar circle. However, APOGEE DR10 suffered from systematic errors in the alpha abundance determinations, particularly for cooler stars. This may have caused errors in their measured abundance gradients, and thus the discrepancy between our results (M. R. Hayden, private communication). The gradient measured for the same stars using APOGEE DR13 is $-0.09 \pm 0.01$ \gradunit which,  although not in agreement with our result, is much more similar (M. R. Hayden, private communication). Gradients measured for APOGEE stars are restricted to $|z| \leq 2$ kpc, which could explain why their measurement is steeper than ours, as we see that the vertical metallicity gradient flattens at larger heights.
 
\subsection{The effects of excluding `transition' stars}
\label{redo}
As mentioned in Section \ref{gmm}, we omitted all stars that lie between the low and high-\afe~populations in terms of abundances and radial velocities so to minimise possible contaminations. In a purely chemical separation, however, they would contribute to the vertical gradients. We explored the effects of excluding them by separating the two populations by [\afe/M] only, making a cut at [\afe/M] = 0.15, where the `gap' is located and repeated our analysis of the gradients. As expected, the vertical metallicity gradients for both sub-populations steepened compared to our probability-based thin/thick disk separation using the [M/H]--[\afe/M]--RV distribution described in Section \ref{gmm}. The low-\afe~population changes to  d[M/H]/d$z=-0.21 \pm 0.01$, and the high-\afe~population to d[M/H]/d$z=-0.11 \pm 0.004$. For the high-\afe~population, we would then be in better agreement with APOGEE DR13 and \cite{Kordopatis2011}. 

\section{[\afe/M] profiles} 
\label{profiles}
\begin{figure}
\centering
		\includegraphics[width=1.05\columnwidth,trim={0.8cm 1cm 0cm 0cm},clip]{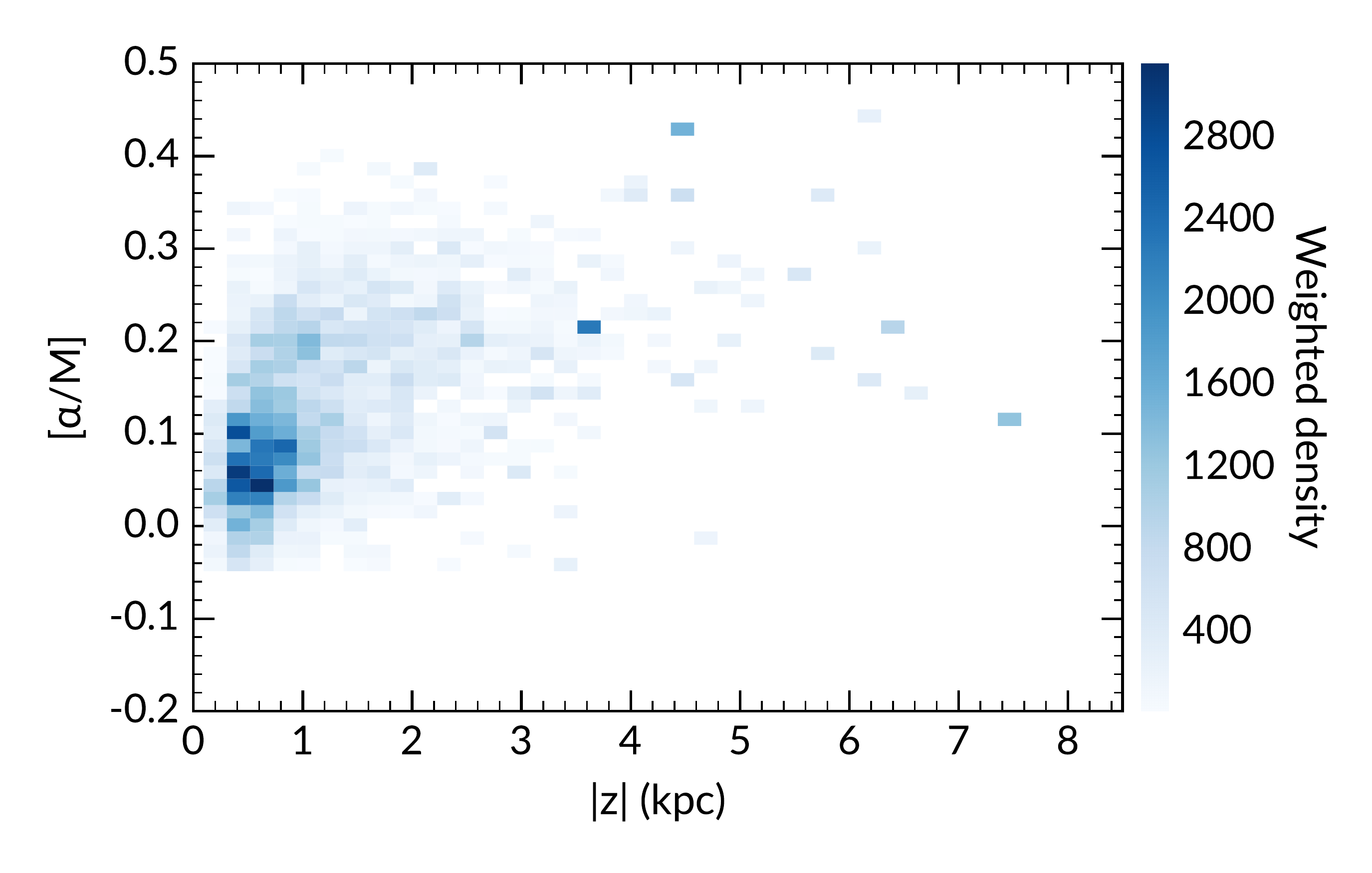}

	\caption{Variation of [\afe/M] with distance from the Galactic plane for all stars. The density is weighted by selection bias fractions as described in Section \ref{bias}. The over densities at high |\emph{z}| are due to a few data points with large weights. Unlike metallicity, [\afe/M] does not vary smoothly with increasing height. There appears to be a break in the distribution at |\textit{z}| $\approx$ 1 kpc.}
	\label{fig:afegrad_all}
\end{figure}

\begin{figure}
\centering
		\includegraphics[width=1.04\columnwidth,trim={0.8cm 1cm 0cm 0cm},clip]{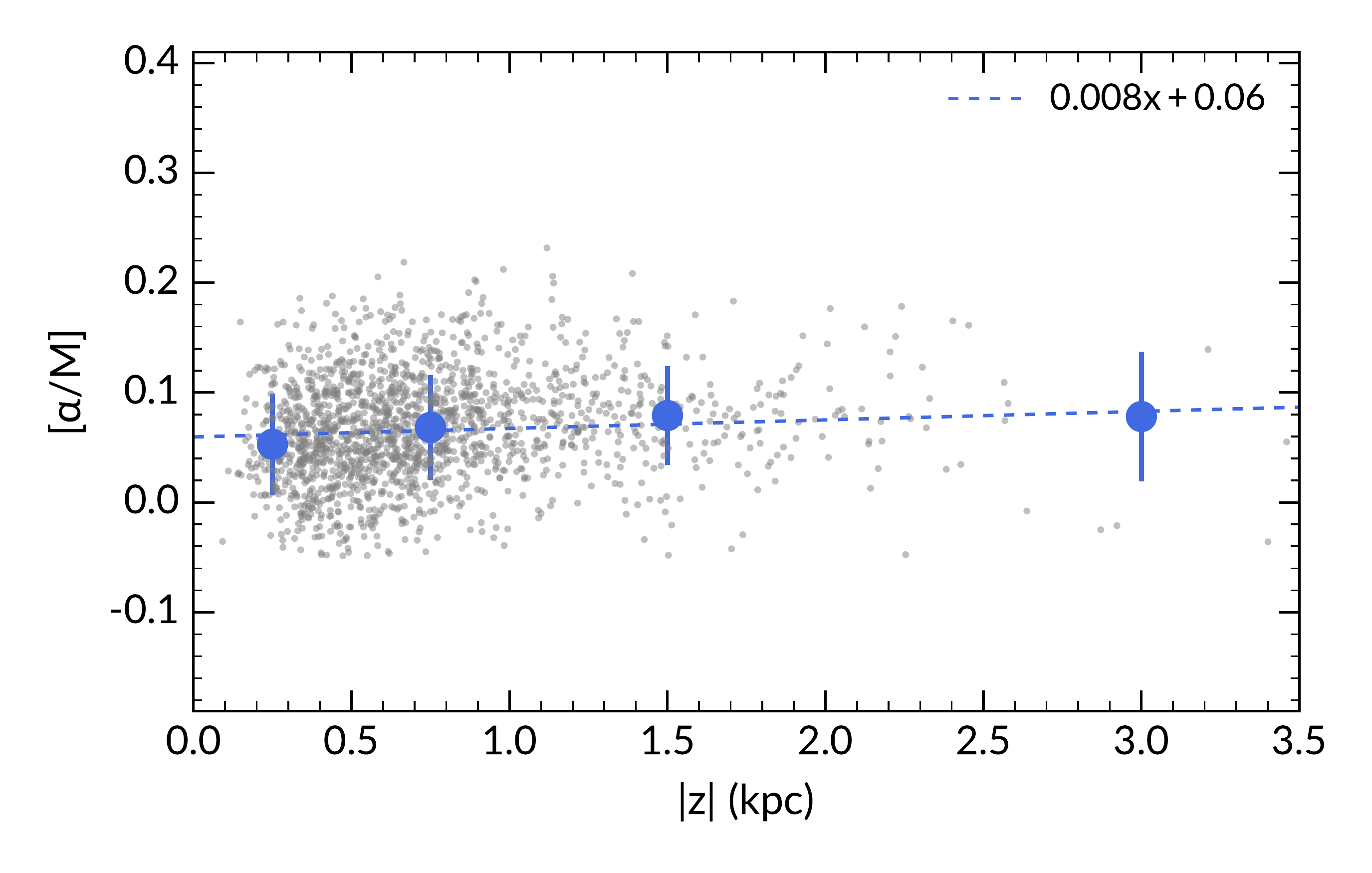}\\
		\includegraphics[width=1.03\columnwidth,trim={0.8cm 1cm 0cm 0cm},clip]{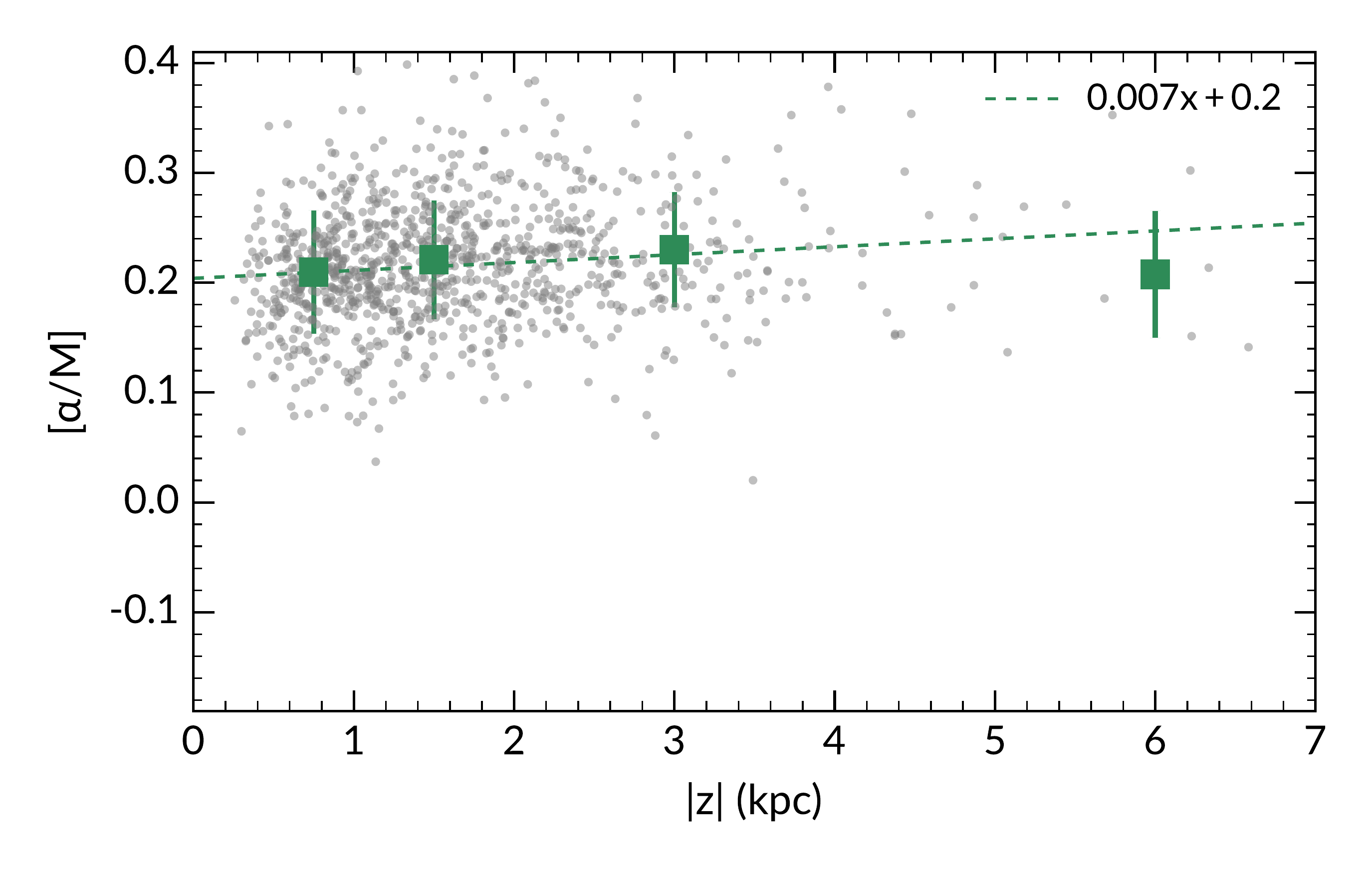}

	\caption{Variation of [\afe/M] with distance from the Galactic plane for each of the high and low-\afe~population. Top panel: vertical abundance gradient for the low-\afe~stars. Bottom panel: vertical abundance gradient of the high-\afe~stars. Both populations show a flat-positive trend. The high-\afe~population shows a higher dispersion in [\afe/M] values. The trends are fitted over grey data points, over-plotted are averaged values of four height bins and their one sigma error bars. Note that the binned values were not used in the gradient fitting.}
	\label{fig:afegrad_thin}
\end{figure}

Within the 1.5 kpc $R_\textrm{GC}$ range of our sample, we do not observe any significant radial changes in [\afe/M] for either of the defined populations. Studies of the high-\afe~population's radial abundance gradients show that there is no variation, but there may be small positive radial [\afe/M] gradients in the low-\afe~population (e.g.,\citealt{Mikolaitis2014,Bergemann2014}). {\cite{Genovali2015}, however, found a negative radial abundance gradient for the $\alpha$-element Ca. \cite{Boeche2014} also found modest radial gradients for the $\alpha$-elements}. The fact that we do not observe a radial abundance gradient in the low-\afe~population is likely due to our limited radial coverage, which prevents us from assessing [\afe/M] variation with $R_\textrm{GC}$.

The vertical \afe-abundance profile of the entire sample is shown in Fig. \ref{fig:afegrad_all}, presented as a density plot similar to Fig. \ref{fig:fehgrad_all}. The median [\afe/M] increases as a function of height, as noted previously by \cite{Schlesinger2014} and \cite{Mikolaitis2014}. However, unlike the metallicity, we find that the \afe-abundance profile does not vary smoothly with $|z|$. 
The \afe-abundance vertical gradient for the entire disk is d[$\alpha$/M]/d$z = 0.038 \pm 0.001$ \gradunit. For the low and high-\afe~populations, the [$\alpha$/M] vertical gradients are both slightly positive, as shown in Fig. \ref{fig:afegrad_thin}. The gradients are d[$\alpha$/M]/d$z = 0.008 \pm 0.002$ \gradunit~for the low-\afe, and d[$\alpha$/M]/d$z = 0.007 \pm 0.002$ \gradunit~for the high-\afe~population. As in Section \ref{redo} above, we also analysed the two \afe~populations with the `transition stars' included. This did not change the slope measured for the low-\afe~population, but increased the slope of the high-\afe~population to d[$\alpha$/M]/d$z$ = $0.014 \pm 0.001$ \gradunit, which is higher than the value measured without transition stars. The transition stars contribute primarily at low $|z|$ ($\leq 1$ kpc), which is why their addition affected the high-\afe~population more: as defined in Section \ref{gmm}, this population is mainly located at $|z| \geq 1$ kpc. 
  
For the high-\afe~population, \cite{Ruchti2011} showed that individual \afe-abundances [Mg/Fe] and [Si/Fe] have vertical gradients $0.03 \pm 0.02$ and $0.02 \pm 0.01$ \gradunit, respectively. Meanwhile, the vertical abundance gradients of [Ca/Fe] and [\ion{Ti}{I,II}/Fe] decrease to $-0.01 \pm 0.01$ and $-0.02 \pm 0.02$ \gradunit. While we do not have \afe-enhanced stars with [M/H] $<-1$ as in their study, this result is in agreement with the flat gradients we observe. \cite{Mikolaitis2014} also provide vertical gradients for the averaged [\afe/M] abundances, as well as vertical gradients for individual \afe-elements using $Gaia$-ESO iDR1. \cite{Mikolaitis2014} found similar vertical abundance profiles for the low and high-\afe~stars. Both populations have averaged and individual vertical \afe-abundance gradients of 0.04--0.05 \gradunit, with errors $<0.01$ \gradunit. These values are not in agreement with our measured vertical gradient for the \afe~sub-populations, as we find that both populations have rather flat abundance distributions as a function of height. However, these results are similar to the gradient we derived for the disk overall. The differences could have arisen from the lack of correction for selection biases in \cite{Mikolaitis2014} and the different abundance scales of the $Gaia$-ESO and GALAH surveys. The [Mg/Fe] histograms shown in Fig. 3 of \cite{Mikolaitis2014} shows that $Gaia$-ESO iDR1 has a larger spread in their abundances compared to GALAH (c.f. Fig \ref{fig:alpha}).

An important point to consider is the dependence of [$\alpha$/M] on [M/H], and the correlation of the latter with respect to vertical height |$z$|. At greater heights above the plane, observed stellar populations become more metal-poor (Fig \ref{fig:fehgrad_all}), and these stars are typically more $\alpha$-enhanced. The positive $\alpha$-gradient over the entire disk is therefore reflective of the fact that more metal-poor, high-\afe~stars become dominant at large heights. For each of the \afe~sub-populations as defined in Fig \ref{fig:gmm}, however, the correlation between [$\alpha$/M] and [M/H] is rather flat, which means that the small positive gradients we measured are intrinsic to these sub-populations. 

\begin{table*}
	\caption{Summary of measured vertical gradients, and intercepts at |\textit{z}| = 0 for disk metallicity and \afe-abundances.}
	\label{table:1}
	\begin{tabular}{lllllllll}
		\hline 
		Population & d[M/H]/d$z$ & $\sigma_{\textrm{d[M/H]/d}z}$ & [M/H]\textsubscript{($z=0$)} & $\sigma_\textrm{[M/H]($z=0$)}$ & d[$\alpha$/M]/d$z$ & $\sigma_{\textrm{d[$\alpha$/M]/d}z}$ & [\afe/M]\textsubscript{($z=0$)} &  $\sigma_\textrm{[\afe/M]($z=0$)}$ \\
		& \gradunit &  & dex &  &\gradunit &  & dex &\\
		\hline
		low-\afe & $-0.18$ & 0.01 & $-0.02$ & 0.01 & +0.008 & 0.002 & 0.06 & 0.002\\
		high-\afe & $-0.058$ & 0.003 & $-0.47$ & 0.01 & +0.007 & 0.002 & 0.20 & 0.003\\
		All stars & $-0.22$ & 0.01 & $-0.08$ & 0.01 & +0.038 & 0.001 & 0.08 & 0.002\\
		\hline
	\end{tabular}
\end{table*}

\section{Discussion} 
\label{discussion}

The process(es) that created the thick disk have been a central point of discussion in Galactic studies. The very definition of the thick disk has changed since first proposed by \cite{gr83}, and here we refer to the `thick disk' as the overall more \afe-enhanced~population as defined in Section \ref{gmm} using both chemical and kinematical information. In Section \ref{intro}, we outlined the main scenarios that have been proposed for thick disk formation, and in this section we interpret our results in the context of these scenarios. 

In summary, the vertical metallicity and abundance profiles of the disk show that:
	
\noindent (1) The disk overall has a steep negative vertical metallicity gradient.

\noindent (2) The low-\afe~population has a similar vertical metallicity gradient to the full disk.

\noindent (3) The high-\afe~population, on the other, has a much flatter vertical metallicity gradient.

\noindent (4) The \afe-abundance ratio increases with height in general. At larger heights only the high-\afe~population is present.  

\noindent (5) Neither the high nor low-\afe~sub-population show a significant vertical alpha abundance ratio gradient.
 
The vertical metallicity gradient in the high-\afe~population is in contrast with predictions of the direct satellite accretion scenario proposed by \cite{aba03} and the fast internal evolution model of \cite{Bournaud2009}. While both of these scenarios could result in a chemically distinct thick disk, they also predict a uniform vertical metallicity distribution, or a lack of vertical metallicity gradient. \cite{Brook2004,Brook2005} proposed that the thick disk formed via merging of gas-rich clumps at high redshift, prior to the formation of the thin disk. Their model predicts an old, \afe-enhanced thick disk that matches observations. However, their thick disk also shows no vertical metallicity gradient, in contrast to our results. 

The heating of an existing disk by small satellite mergers can create a thick-disk like vertical structure (e.g., \citealt{Quinn1986,Quinn1993,Kazantzidis2008,Villalobos2008}). {The vertical metallicity gradient of the existing disk could be preserved in the thick disk, however this can also be affected by the interplay between radial migration and radial metallicity gradients (e.g.,~\citealt{Kawata2018})}. \cite{Bekki2011} modelled the kinematics and chemistry of stars formed by minor mergers in detail, and showed that a fast star formation rate in the thick disk results in \afe-enhanced stars. The steep vertical metallicity gradient of the pre-existing disk flattens over time, but qualitatively it is steeper in the inner Galaxy, consistent with the observations of \cite{Hayden2014}. However, the final thick disk gradient is essentially flat at the solar circle, which is not what we observe. Furthermore, disk flaring is expected in such a heating scenario. For the high-\afe~stars, \cite{Bovy2016} did not observe any flaring in their mono-abundance populations (MAPs). However, \cite{Minchev2015,Minchev2016} argued that mono-age populations always flare in their cosmological simulations, and MAPs (mono-abundance populations) are not necessarily co-eval. Based on APOGEE abundances and calibrated ages\footnote{Ages from \cite{Martig2016}, calibrated on APOGEE DR12 C and N abundance ratios.},~\cite{Mackereth2017} found that mono-age \afe-enhanced populations do show some flaring, albeit with a smaller amplitude compared to the low-\afe~population. Further observational and model constraints from stellar ages and flaring of the high-\afe~stars are thus needed to understand the importance of minor mergers in thick disk formation. 

The secular radial migration 
\citep{selbin02} process was proposed by \cite{Schonrich2009} as the sole explanation for the thick disk {(but this remains controversial, see e.g., \citealt{Minchev2010,Vera-Ciro2014})}. Stars in the inner galaxy are formed fast and migrated outwards to create the \afe-enhanced population at large scale heights. Since radial migration is more likely to affect older stars, a negative vertical metallicity gradient, and a positive [\afe/M] gradient are expected. \cite{Schonrich2017} obtain d[M/H]/d$z \approx -0.2$ \gradunit~for the full nearby disk in their analytical model (which included inside-out disk formation), in agreement with our observations. \cite{Loebman2011} also reported a similar vertical gradient of $\approx$$-0.18$ \gradunit~in their N-body simulation with extensive radial migration. Both of these values are consistent with our measurements of the full disk and low-\afe~population, even though \cite{Loebman2011} did not calibrate their model to reproduce the Milky Way. 

Radial migration signatures are observed in the metallicity distribution function (MDF) of disk stars at different Galactocentric radii. \cite{Hayden2015} observed that at small heights above the plane, the skewness of MDFs changes from negative in the outer galaxy (skewed towards metal-poor stars) to positive (skewed towards metal rich stars) in the inner galaxy. In contrast, the high-\afe~population's MDF remain constant at all locations. \cite{Loebman2016} showed that these observations can be qualitatively explained by radial migration in their simulation. The change in skewness of the disk MDFs at different radii could be due to an increased fraction of migrated stars beyond $\approx$5 kpc, such that more metal rich stars are migrated to larger $R_\textrm{GC}$. As the high-\afe~stars formed within a small region and a few Gyrs in a well mixed environment, their chemical content are similar and thus the MDF remains constant at all Galactic locations. \cite{Loebman2016} found a small vertical metallicity gradient in their simulated high-\afe~population of $\approx$$-0.03$ \gradunit. This is half the value observed in our study. 
Further investigation of the vertical metallicity and abundance gradients for the high-\afe~stars in radial migration models will help to determine the extent to which it affects this population.  

\cite{Bird2013} was able to produce a Milky Way-like galaxy with an old, vertically extended population much like the Galactic thick disk using the ``Eris'' cosmological simulation suite. The effects of the active merger phase at early times (redshift $>$ 3), secular heating, and radial migration on the present-day galaxy were examined. \cite{Bird2013} found that stars born during the merger phase have larger scale heights and shorter scale lengths, and younger populations form progressive thinner and longer structures. This gradual transition from a kinematically hot and thick disk to a colder, thinner disk was dubbed `upside down' formation (see also \citealt{Samland2003}). Interestingly, secular heating and radial migration did not have a large impact on the final properties of each coeval population. Rather, the trends are established at formation, suggesting that the thick disk-like component was born thick. Similarly, \cite{Stinson2013} and \cite{Brook2012} concluded that their \afe-enhanced, older populations were born kinematically hot, and that the early disk settles into a thin component. The settling process of the galaxies and fast formation of the old, \afe-enhanced and vertically extended populations in these simulations could produce the vertical metallicity and abundance profiles observed in this work. It was shown by \cite{Wisnioski2015} that the observed velocity dispersion of H$\alpha$ gas in galaxies at high redshift decreases with time, providing further indication that disk galaxies were born thick at redshifts of $z = 1-2$. 

While the cosmological models mentioned above heavily rely on the condition that disk galaxies like the Milky Way had a quiescent merger history, there is observational evidence that this may be true for the Galaxy \citep{Ruchti2015,Casagrande2016}. However, the metallicity and \afe-abundance gradients in these simulations have not been studied in detail. 

 Although we observe an overall continuity in the vertical metallicity profile, we see two distinct \afe-enhancement tracks as a function of $|z|$, which have implications for the star formation history of the disk. \cite{Haywood2013,Haywood2016} proposed two different star formation epochs for the high and low-\afe~stars. By comparing their chemical evolution model \citep{Snaith2015} with APOGEE data, \cite{Haywood2016} proposed that the star formation rate dropped significantly at ages of 10 Gyr before increasing again at about 7 Gyr to a lower maximum value. This could indicate the transition between thick to thin disk formation. However, the authors note that due to the strict continuity of the stellar abundances, the gas supply must not have decreased during this period of time. Similarly, \cite{Brook2012} found that the star formation rate decreased slightly at around 7 Gyr, near the epoch of thin disk formation in their simulation. {This idea was also explored by the two-infall model proposed by \cite{Chiappini1997}, who argued for a decreased star formation rate between the epochs of halo-thick disk and thin disk formation, and proposed a shorter formation timescale for the halo/thick disk of 1 Gyr.} Future work that incorporate stellar ages (e.g., from the GALAH/K2 overlap) will be able to rigorously test these scenarios {and provide additional constraints on the formation time-scale of the thick disk}. 
 
\section{Conclusion} 
\label{conclusions}

We have determined the vertical profiles of metallicity and \afe-abundances in the Galactic disk using data from the GALAH first internal data release. We analysed in total 3191 giants from the GALAH pilot and main surveys, extending up to 4 kpc in height above the plane, within a small range of Galactocentric distance ($7.9 \la R_\mathrm{GC} \la 9.5$ kpc). The precise metallicity and abundance measurements of GALAH allow us to reliably define `thick' and `thin' disk populations using chemistry and radial velocities. The GALAH magnitude limits in the estimated $V$-band translate to a dependency in $(J-K)$ colour and magnitudes. We corrected for the selection effects for targets from the pilot and main surveys separately by population synthesis using BaSTI isochrones. 

The vertical metallicity gradient of the entire disk is $-0.22 \pm 0.01$ \gradunit, which is in agreement with recent estimates from large spectroscopic surveys such as SEGUE and APOGEE. The low-\afe~population, or the thin disk, also exhibits a steep negative vertical metallicity gradient d[M/H]/d$z=-0.18 \pm 0.01$ \gradunit. The more enhanced \afe~population, which we identify as the thick disk, is found to have a shallower vertical gradient d[M/H]/d$z$ of $-0.058 \pm 0.003$ \gradunit. { We note again that our data do not probe the metal-poor extension of the thick disk}, however, the vertical gradients observed here are similar in amplitude to those of previous studies. Overall, our results confirm some conclusions reached by earlier studies, despite differences in target selection, spatial coverage and abundance scales. 
The discrepancies were likely caused by uncorrected selection effects in some cases, and the many different definitions in the literature of high-\afe, or thick disk stars. 

As expected, [\afe/M] increases as a function of $|z|$, with the low-\afe~population occupying lower heights on average. The vertical [\afe/M] profile at the solar circle shows that there are two over-densities, with the discontinuity most clearly seen around $|z| = 1$ kpc. We find that the both low and high-\afe~sub-populations have a flat vertical [\afe/M] gradient. Similarly, \cite{Ruchti2011} also found flat vertical gradients for individual \afe-abundances at the metal-poor end of the \afe-enhanced population. For the low-\afe~population the gradient can be explained by radial migration playing an important role in the evolution of the thin disk. The negative vertical metallicity gradient in the high-\afe~population indicate that scenarios which have uniform `thick disk' vertical metallicity gradients are not responsible for its formation. The vertical [M/H] gradient observed in this work and elsewhere could have arisen from a settling phase of the disk as suggested by \cite{Samland2003} and \cite{Bird2013}, minor heating episodes such as in the models of \cite{Kazantzidis2008,Villalobos2008}, or caused by radial migration \citep{Schonrich2009,Loebman2011}. Mergers cause flaring of the disk, which is seen in the low-\afe~population in the analysis of \cite{Bovy2016}, but not in the high-\afe~population. However, \cite{Mackereth2017} have since shown that coeval high-\afe~populations do indeed show flaring, but much less than the low-\afe~stars. On the other hand, the \afe-abundances of both sub-populations are distinct and nearly constant at all heights, indicating that they are formed in very different conditions. 

Accurate distances and proper motion from \emph{Gaia} DR2 will allow for an even more accurate and detailed analysis of the chemistry and kinematics of the high-\afe~population, not only for the GALAH pilot survey but also the larger GALAH main sample. This will give us a clearer and more definitive picture of the formation and evolution of the Milky Way thick disk.

\section*{Acknowledgements}
We thank Michael R. Hayden for helpful discussions regarding APOGEE results, and the anonymous referee for comments that improved the clarity of this manuscript. LD and MA acknowledge funding from the Australian Government through ARC Laureate Fellowship FL110100012. LD, KCF and RFGW acknowledge support from ARC grant DP160103747. LC gratefully acknowledges support from the Australian Research Council (grants DP150100250, FT160100402). DMN was supported by the Allan C. and Dorothy H. Davis Fellowship. DS is the recipient of an Australian Research Council Future Fellowship (project number FT1400147). TZ acknowledges financial support from the Slovenian Research Agency (research core funding No. P1-0188). Part of this research was supported by the Munich Institute for Astro- and Particle Physics (MIAPP) of the DFG cluster of excellence "Origin and Structure of the Universe". 

This publication makes use of data products from the Two Micron All Sky Survey, which is a joint project of the University of Massachusetts and the Infrared Processing and Analysis Center/California Institute of Technology, funded by the National Aeronautics and Space Administration and the National Science Foundation; and data products from the Wide-field Infrared Survey Explorer, which is a joint project of the University of California, Los Angeles, and the Jet Propulsion Laboratory/California Institute of Technology, funded by the National Aeronautics and Space Administration.

\bibliographystyle{mnras}
\bibliography{Ref} 

\appendix
\section{Online data table}
\begin{table*}
	\caption{An example of the contents included in the online data table. The complete table is available on the publisher website.}
	\label{table:a1}
	\begin{tabular}{cccccc}
		\hline 
		OBJECT ID & UCAC4 ID & RAJ2000 & DEJ2000 & Distance & Pr(thick)  \\
		&  &  deg  & deg &pc &  \\
		\hline
131216002101010 &156-011705& 115.5245& -58.9095& 1102.72& 0.1583\\
131216002101012& 156-011845& 116.1019& -58.9680& 1502.33& 0.0009\\
131216002101013& 156-011859& 116.1602& -58.9823& 3058.82& 0.0188\\
131216002101018& 156-011814& 115.9660& -58.9922& 854.35& 0.0083\\
131216002101021& 155-011860& 116.3630& -59.0953& 2287.51& 0.0180\\
131216002101023& 155-011759& 116.0724& -59.0578& 2200.28& 0.0096\\
131216002101025& 155-011806& 116.2236& -59.1011& 2078.00& 0.2680\\
131216002101028& 156-011737& 115.6524& -58.9935& 628.69& 0.0003\\
131216002101029& 155-011714& 115.9342& -59.0706& 1578.78& 0.0079\\
131216002101033& 155-011844& 116.3374& -59.1984& 871.58& 0.0021\\
		\hline
	\end{tabular}
\end{table*} 

We have included a data table listing the stars analysed in this work, their GALAH object ID, UCAC4 catalogue ID, coordinates, thick disk membership probability and distances. Table \ref{table:a1} provides an example of the contents included in the online material. 

\label{lastpage}
\end{document}